\documentclass[aps,pre,twocolumn,amsmath,amssymb,amsfonts]{revtex4}
\usepackage{epsfig}
\usepackage{graphicx}
\usepackage{color,soul}
\usepackage{dcolumn}

\usepackage{bm}\usepackage[titletoc]{appendix}
\begin{document}
\title{Limit theorems for L\'{e}vy walks in $d$ dimensions: rare and bulk fluctuations \footnote{Submitted to Smoluchowski's  special issue Gudowska-Novak, Lindenberg, Metzler, Editors.} }
\author{Itzhak Fouxon$^{1,2}$}\email{itzhak8@gmail.com}
\author{Sergey Denisov$^{3,4}$}\email{sergey.denisov@physik.uni-augsburg.de}
\author{Vasily Zaburdaev$^{3,5}$}\email{vzaburd@pks.mpg.de}
\author{Eli Barkai$^{1}$}\email{eli.barkai@biu.ac.il}
\affiliation{$^1$ Department of Physics, Institute of Nanotechnology and Advanced Materials, Bar-Ilan University, Ramat-Gan, 52900, Israel}
\affiliation{$^2$ Department of Computational Science and Engineering, Yonsei University, Seoul 120-749, South Korea}
\affiliation{$^3$ 
Institute of Supercomputing Technologies, Lobachevsky State University of Nizhny Novgorod, Gagarina Av. 23, 
Nizhny Novgorod, 603140, Russia}
\affiliation{$^4$ Sumy State University, Rimsky-Korsakov Street 2, 40007 Sumy, Ukraine}
\affiliation{$^5$ Max Planck Institute for the Physics of Complex Systems, N\"{o}thnitzer Strasse 38, D-01187 Dresden, Germany}

\begin{abstract}
We consider super-diffusive L\'{e}vy walks in $d \geqslant 2$ dimensions when the duration of a single step, i.e., a ballistic motion performed by a walker, is governed by a power-law tailed distribution of infinite variance and finite mean. We demonstrate that the probability density function (PDF) of the coordinate of the random walker has two different scaling limits at large times. One limit describes the bulk of the PDF. It is the $d-$dimensional generalization of the one-dimensional L\'{e}vy distribution and is the counterpart of central limit theorem (CLT) for random walks with finite dispersion. In contrast with the one-dimensional L\'{e}vy distribution and the CLT this distribution does not have universal shape. The PDF reflects anisotropy of the single-step statistics however large the time is. The other scaling limit, the so-called 'infinite density', describes the tail of the PDF which determines second (dispersion) and higher moments of the PDF. This limit repeats the angular structure of PDF of velocity in one step. Typical realization of the walk consists of anomalous diffusive motion (described by anisotropic $d-$dimensional L\'{e}vy distribution) intermitted by long ballistic flights (described by infinite density). The long flights are rare but due to them the coordinate increases so much that their contribution determines the dispersion. We illustrate the concept by considering  two  types of  L\'{e}vy walks, with isotropic and anisotropic distributions of velocities. Furthermore, we show that for isotropic but otherwise arbitrary velocity distribution the $d-$dimensional process can be reduced to one-dimensional L\'{e}vy walk.

\end{abstract}  \maketitle

\section{Introduction}

It is a universal consequence of microscopic chaos that 
the velocity $\bm v(t)$ of a moving particle has a finite correlation time \cite{Sinai,bunim}. 
The particle's displacement $\int_0^t \bm v(t')dt'$ 
on large time scales can be considered as an outcome of a sum of independent random 'steps'. 
A single step here is a motion during the shortest time  over which the velocity of the particle is correlated. 
Velocity correlations at different steps can be neglected and the integral over the time interval $[0,t]$ 
can be replaced with the sum of integrals over disjoint steps. 
If variations of step durations and  velocity fluctuations
can be neglected as well,  we arrive at the standard random walk 
consisting of steps that  take the same fixed time. 
The distance covered during single step is fixed  but the direction of the step is random. 
It can be, for example,  a step along one of the basis vectors  when random walk is performed  on a $d-$dimensional lattice or 
it can be a step in a completely random direction in $d-$dimensional space as in the case of isotropic walk. In many situations though 
the variation of the step's duration and 
velocity cannot be disregarded. A famous example is L\'{e}vy 
walks (LWs) \cite{mw,sl,Shl,Mon,Shlesinger82,bg,Isichenko,Shlesinger93,Shlesinger96,UchZol,km,godreche,ks,xper,BarkaiPhysRev,
BarkaiPhysLett,BarkaiPR,Zaburdaev,Marcin,recent,arxiv,kb,Burioni,Dentz,Agliari} 
that belong to a more general class of stochastic processes called continuous time random walks (CTRWs)~\cite{mw,sl}.
In CTRWs it is not only the direction of the displacement during one step that is random (as in the standard random walk) 
but also the length of the step and the time that it takes. 
This flexibility allows to cover a large number of real-life situations including dynamics of an ordinary gas molecule when both the time between 
consecutive collisions and velocity of the molecule vary. In ideal gas the probability of large (much larger than the mean free time) 
time between the collisions is negligibly small 
so the probability density function (PDF) 
of the step duration decays fast for large arguments. 
This is not the case for the so-called Lorentz billiard \cite{bunim}, in which
the particle moves freely between collisions with scatterers arranged in a spatially periodic array. 
In the case without horizon, when infinitely long corridors between the scatterers are present, the particle 
can fly freely for a very long time if its velocity vector aligns close to the direction of the corridor. 
The distribution of times between consecutive collisions has a power-law tail
and  infinite variance \cite{bd}.  The dynamics of a particle can be reproduced with a L\'{e}vy walk process to great detail \cite{cristadoro,den1}.
L\'{e}vy walks were found
in diverse real-life processes including the spreading of cold atoms in optical lattices, 
animal foraging, and diffusion of light in disordered  glasses and hot atomic vapors (for more examples see a recent review \cite{Zaburdaev} and references therein). 
Despite of these advances and new experimental findings, the theory of LWs remains mainly confined to the case of one-dimensional geometry.

In this work we study  $d-$dimensional L\'{e}vy walks
when the duration $\tau$ of single steps is characterized by a PDF $\psi(\tau) $ with power-law asymptotic form, $\psi(\tau) \varpropto \tau^{-1-\alpha}$. 
The most interesting and practically relevant is the so-called 'sub-ballistic super-diffusive regime' \cite{Zaburdaev}, $1 < \alpha  < 2$,  when
the step time has finite mean but infinite variance \cite{note1}. We study the PDF $P(\bm x,t)$ of the walker's coordinate $\bm x(t)$ at time $t$ in this regime. 
We show that there are two ways of rescaling $P(\bm x, t)$ with powers of time that produce finite infinite time limits. The two limiting distributions describe the bulk and the tails of $P(\bm x, t)$. Both distributions are sensitive to the microscopic statistics of the velocity of walkers and are model specific. 

We consider a random walk in $\mathbb{R}^d$ with a PDF $F(\bm v)$ of the single step velocity that obeys $F(-\bm v)=F(\bm v)$, cf. \cite{antipod,direction}. Thus the process is unbiased and  the average displacement is zero.
In the case of the sum of a large number of independent and identically-distributed (i. i. d.)
random variables with finite dispersion there is a well-known scaling, 
\begin{eqnarray}&&\!\!\!\!\!\!\!\!\!\!\!\!\!\!\!
\lim_{t\to\infty}t^{d/2}P\left(\bm x\sqrt{t}, t\right)=g(\bm x), \label{g}
\end{eqnarray}
where $g(\bm x)$ is a Gaussian distribution \cite{Feller}. 
A stronger, large deviation limit tells that 
\begin{eqnarray}&&\!\!\!\!\!\!\!\!\!\!\!\!\!\!\!
\lim_{t\to\infty}\frac{1}{t} \ln P(t\bm x, t)=s(\bm x),\label{largedeviations}
\end{eqnarray}
where the convex function $s(\bm x)$ is known as 
large deviations, or entropy, or rate, or Kramer's function \cite{ellis1,ellis2}; see \cite{fb} for simple derivation. 
This limit describes exponential decay of the probability of large deviations of finite-time value $\bm x(t)/t$ 
from its infinite time limit fixed by the law of large numbers, $\lim_{t\to\infty} (1/t)\bm x(t)=0$. 
It corresponds to the Boltzmann formula and tells that the PDF of macroscopic thermodynamic variable (representable as the sum of large number of 
independent random variables) is exponential of the entropy whose maximum's location gives the average. The coefficients of the quadratic expansion near the maximum characterize 
thermodynamic fluctuations \cite{ellis2}. 
Thus in the case of finite dispersion there is one universal scaling with 
the limiting distribution (the special case of finite dispersion but 
power-law tail, $\alpha >2$,  produces different limiting distribution \cite{Rebenshtok}; we do not consider this case here).

\begin{figure*}[t]
\includegraphics[width=1.\textwidth]{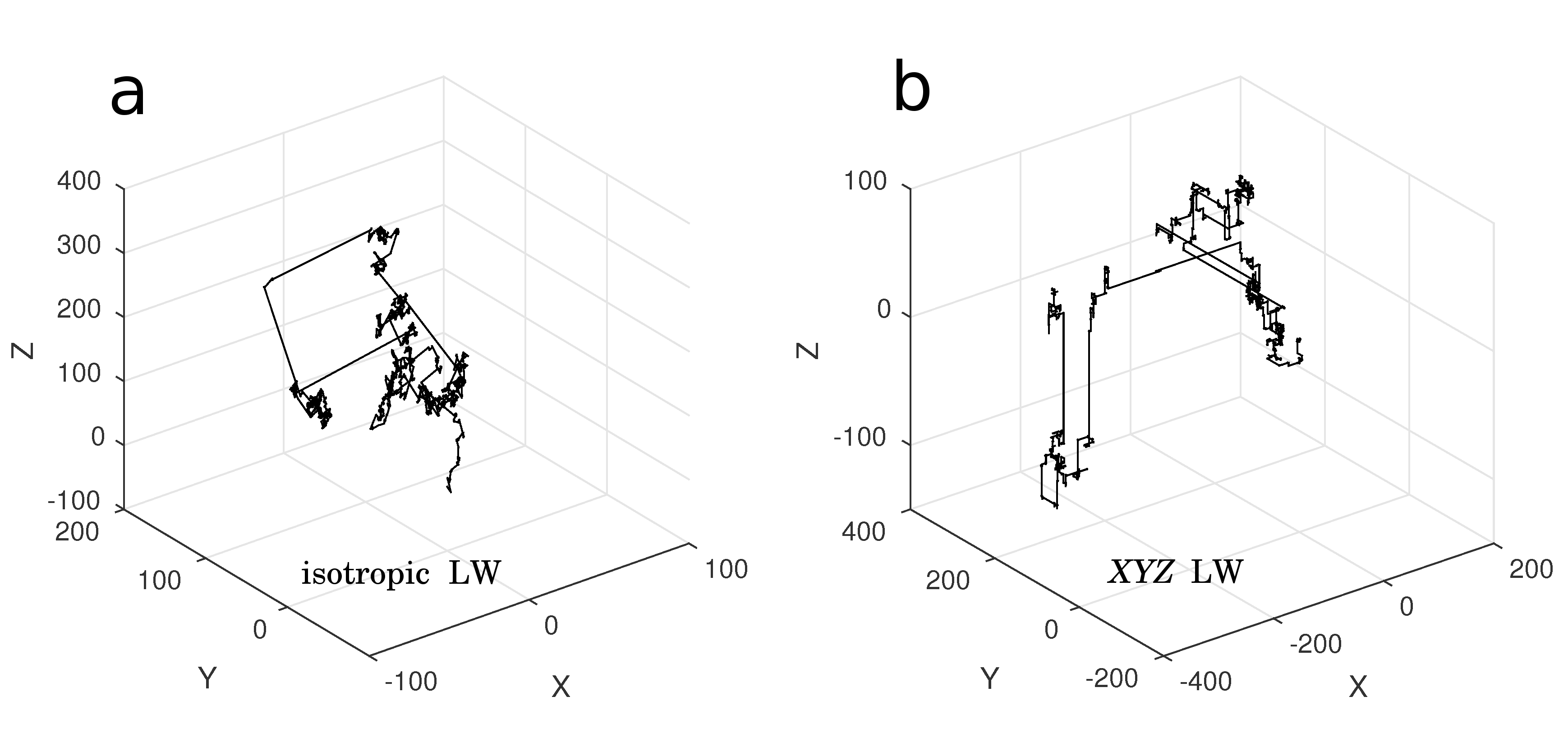}
\caption{\textbf{Examples of  L\'{e}vy walks in three dimensions.} A walker has
a constant (by absolute value) velocity $\upsilon_0 = |\bm v| =1$. 
When performing \textit{uniform} L\'{e}vy walk (a), the walker, after completing a ballistic flight, instantaneously selects a random time $\tau$ for a new  flight
and randomly chooses a flight direction (it can be specified by a point on the surface of the unit sphere).
The velocity PDF $F(\bm v)$ is described by the uniform distribution over the unit sphere's surface.  
In the  case  of anisotropic \textit{$XYZ$} L\'{e}vy walk (b), the  walker is allowed 
to move only along one axis at a time. After the completion of a flight, the walker 
selects a random time $\tau$ for a new flight, 
one out of six directions (with equal probability)  and then moves along chosen direction. The velocity PDF $F(\bm v)$ in this 
case is six delta-like distributions located at the points where the unit sphere is penetrated by the frame axes.
The parameter $\alpha$ is $3/2$. In our simulations we used $\psi(\tau) = (3/2) \tau^{-5/2}$ for $\tau>1$ otherwise it is zero.}
\label{Fig1}
\end{figure*}

There are two different scalings that were found in the case of one-dimensional super-diffusive LWs \cite{BarkaiPhysRev,BarkaiPhysLett}.
One of them is a continuation of the central limit theorem (CLT) to the case of i. i. d. random variables with infinite dispersion, the so-called 
generalized central limit theorem (gCLT) \cite{lv}. 
The corresponding distribution  is known as the celebrated L\'{e}vy distribution. We first generalize this scaling to the $d-$dimensional case
and find significant difference from the one-dimensional 
case and ordinary $d-$dimensional random walks. Our result shows that for $d-$dimensional L\'{e}vy walks with power law tailed distribution of step duration $\tau$ that has infinite moments starting from order $\alpha$ between $1$ and $2$ there is finite limit, 
\begin{eqnarray}&&\!\!\!\!\!\!\!\!\!\!\!\!\!\!
\lim_{t\to\infty} t^{d/\alpha}P(t^{1/\alpha}\bm x, t)\nonumber\\&&\!\!\!\!\!\!\!\!\!\!\!\!\!\!=\int \exp\left[i\bm k\cdot\bm x\!-\!\frac{A}{\langle \tau\rangle}\left|\cos\left(\frac{\pi\alpha}{2}\right)\right|\left\langle |\bm k\cdot\bm v|^{\alpha}\right\rangle \right]\frac{d\bm k}{(2\pi)^d}.\label{g0}
\end{eqnarray}
This holds for arbitrary statistics of particle's velocity that has finite moments and obeys $F(\bm v)=F(-\bm v)$. We demonstrate that this result smoothly connects with the usual central limit theorem given by Eq.~(\ref{g}). Indeed Eq.~(\ref{g0}) reproduces Gaussian distribution at $\alpha=2$. Thus our result includes the usual central limit theorem as particular case and can be called generalized central limit theorem. We clarify in Sec. \ref{sm} that Eq.~(\ref{g0}) can be obtained from distribution of sum of many independent identically distributed scalar random variables with power-law tailed distributions, cf. \cite{levy,lv,bg,UchZol,kg,st}.

The limiting distribution given by Eq.~(\ref{g0}) has features that distinguish it from both the one-dimensional LW and $d-$dimensional random walks. In those cases there is universality: 
the form of the scale-invariant PDF does not depend on details of the single step statistics 
(for instance, ordinary two-dimensional random walks on triangular and square lattices are described by the same isotropic Gaussian
PDF in the limit of large times \cite{Itzykson}). 
In the case of $d-$dimensional LWs  the anisotropy of the single step statistics  is imprinted into 
the statistics of the displacement -- no matter how large the observation time $t$ is; see the discussion of the particular case of $d = 2$ in  Ref.~\cite{recent}. 

In the language of field theory, ordinary random walks are \textit{renormalizable}:
in the long-time limit the information on macroscopic structure of 
the walks is reduced to finite number of constants \cite{Itzykson}. 
Thus the Gaussian PDF of the sum of large number of i. i. d.
random variables is fully determined  by the mean and the dispersion of those variables and their total number. 
The rest of the information on the statistics of these variables is irrelevant -- when the Gaussian bulk of the PDF is addressed. 
In contrast, the entropy function of the large deviations theory is the Legendre transform of the logarithm of the characteristic function 
of the random variable in the sum. Thus large deviations description is sensitive 
to the details of statistics of one step of the walk. 
We  conclude that the Gaussian bulk of the PDF of the sum of large number of i. i. d. random variables 
with finite dispersion is determined by a finite number of constants characterizing the statistics of the single step but the tail of the PDF is not. 

In this work we show that the PDF of a $d$ - dimensional
LW is not universal and   cannot be specified with a finite number of 
constants -- already in the bulk -- if the statistics of the velocity of single steps, given by the 
PDF $F(\bm v)$, is anisotropic. To stress this fact we call the corresponding  distributions described by Eq.~(\ref{g0})
'anisotropic L\'{e}vy distributions'. In contrast,  if the velocity statistics is 
isotropic, for example, $F(\bm v)$ is given by the uniform distribution on the surface of the $d$-dimensional sphere of the radius $\upsilon_0 = const$ (see Fig.~1a),
the description of the process  can be reduced to one-dimensional case and the bulk of the 
corresponding PDF is fully determined  by a finite number of constants which can be calculated from 
$F(\bm v)$. Thus in the anisotropic case there is a dramatic difference between 
the bulk of the PDFs of a LW and the random walk with finite dispersion of the single step duration. 

We demonstrate that besides the limiting distribution given by Eq.~(\ref{g0}) there is another limiting distribution which is determined by 
\begin{eqnarray}&&\!\!\!\!\!\!\!\!\!\!\!\!\!\!
\lim_{t\to\infty}\!\frac{P(t \bm v, t)}{t^{1-d-\alpha}}\!=\frac{A}{v^{d-1}|\Gamma(1-\alpha)|\langle \tau\rangle}\nonumber\\&&\!\!\!\!\!\!\!\!\!\!\!\!\!\!\times\int_{v'>v} F(v'{\hat v})v'^{d-1}d v' \left[\alpha\frac{v'^{\alpha}}{v^{1+\alpha}}-(\alpha-1)\frac{v'^{\alpha-1}}{v^{\alpha}}\right], 
\end{eqnarray}
where $\bm v=\bm x/t$ is the effective velocity of the particle, $\bm v=v{\hat v}$ and the limit exists because the tail of the PDF is determined by ballistic-type events. This distribution is called infinite density where the word 'infinite' refers to the non-normalizable character of this function found previously in one-dimensional case \cite{aa,tz,BarkaiPhysRev,BarkaiPhysLett}. This pointwise limit holds for $\bm v\neq 0$ non-contradicting normalization of the PDF: in this limiting procedure the normalization is carried by $v=0$ point. 
The existence of the other scaling limit is unique property that has origins in the scale-invariance of the tail of distribution of $\tau$. This distribution describes the tail of the PDF of the particle's coordinate. In this sense it is the counterpart of the large deviations result for ordinary random walk given by Eq.~(\ref{largedeviations}). There is however significant difference: the large deviations function describes averages of high-order moments but in the case of L\'{e}vy walks the infinite density provides already dispersion of the process. This can be seen observing that as we demonstrate the distribution provided by Eq.~(\ref{g0}) has power-law tail with divergent second moment.
This is because for L\'{e}vy walks rare events when the walker performs extremely long ballistic flights have a substantial impact on the total displacement of the walker 
even in the limit of long times. The probability of long ballistic steps is not negligibly small and single ballistic steps could be discerned in a single trajectory of the walker for any time $t$. Such steps form the outer regions of the PDF.

Thus the two limit distributions describe the bulk and the tail of the LW's PDF, respectively. 
The bulk is formed by the accumulation of typical (most probable) steps. 
They are responsible for a diffusive motion (albeit already anomalous one). 
In contrast, the PDF's tails are
formed  by long ballistic flights and they are described by the infinite density. These flights are rare steps where the walker moves 
for a long (i. e., comparable  to the total observation time $t$) time without changing its velocity. 
In the case of the Lorentz billiard this is the situation when the velocity vector of the particle
aligns close to the direction of one of the ballistic corridors \cite{cristadoro}. 
Though this happens relatively rarely,
the distance covered by the walker during a such flight is so large that these flights give \textit{finite} contribution to
the probability that the walker displacement after the time $t$  is of the order $v_0t$.
When the probability of long flights is, for example,  exponentially small (as in the case of the standard random walks)
the contribution of the flights can be neglected. This is not, however, the case of the LWs with power-law asymptotic of $\psi(\tau)$, 
as we demonstrate in this paper.  

The paper is organized as follows. In Section \ref{Form} we introduce the basic definitions and the tool of the study - the Fourier-Laplace transform of the PDF of the walker's coordinate. In the next Section we provide complete solution for the case of isotropic statistics of velocity of the walker. Central result of our work - the anisotropic CLT for the bulk of the PDF is derived in Section \ref{CLT}. The next Section describes universality of the tail of this non-universal bulk that helps finding low moments of the distance from the origin. Section \ref{infn} provides the other limiting theorem on infinite density that provides the tail of the distribution. The next Section provides detailed form of the moments of arbitrary order including anomaly in growth due to anisotropy. In Section \ref{frct} we provide the tail of the PDF and Conclusions resume our work.

\section{Fourier-Laplace transform of the PDF $P(\bm x,t)$}\label{Form}

In this Section we specify the considered random process and introduce the main tool of the analysis on which all further results rest. This is the Fourier (in space) - Laplace transform (in time) of $P(\bm x,t)$.

We consider a LW as an infinite sequence of flights (steps) of random duration $\tau_i$ 
where $i$ is the flight's index in the chronologically ordered sequence. 
The process starts at time $t=0$ at the point  $\bm x=0$. 
The velocity of the walker during a flight is a random  vector $\bm v_i$ which remains constant during the flight. 
Upon the completion of the flight 
both the velocity $\bm v_{i+1} $ and the duration $\tau_{i+1}$ of the next flight are randomly chosen, by using PDFs $F(\bm v)$ and $\psi(\tau)$, respectively.
The number of flights performed during the observation time $t$, $N(t)$, is a random number constrained by
$t = \sum_{i=1}^{N(t)} \tau_i + \tau_b\ $, where $\tau_b = t-t_N$ is so-called the backward recurrence
time \cite{godreche}. The time $t$ coordinate of the particle is,  
\begin{eqnarray}&&\!\!\!\!\!\!\!\!\!\!\!\!\!\!
\bm x(t)=\sum_{i=1}^{N(t)}\bm v_i\tau_i+\bm v_{N(t)+1}\left(t-\sum_{i=1}^{N(t)}\tau_i\right). \label{walkoro}
\end{eqnarray}
The simplest model is  $d-$dimensional 'L\'{e}vy plotter', the product 
of $d$ independent one-dimensional walks along the basis vectors which span $\mathbb{R}^d$. The PDF of this process is the product 
of the corresponding one-dimensional PDFs  \cite{BarkaiPhysRev,BarkaiPhysLett,Zaburdaev}. 
This case demands no further calculations so we next consider non-trivial set-ups.

We consider two intuitive models, the uniform LW and anisotropic $XYZ...$ LW. 
In the \textit{uniform model} $F(\bm v)$ is specified by the uniform distribution on the surface of $d-$dimensional unit sphere so velocity has fixed magnitude $1$; see Fig.~1a. As we demonstrate in the next section,  in many respects this model can be reduced to the one-dimensional case. 
In the \textit{anisotropic $XYZ...$} model particle moves along one of the $d$  basis vectors at a time; see Fig.~1b. 
The analysis we present  below is valid for any PDF $F(\bm v)$ obeying 
the symmetry  $F(-\bm v)=F(\bm v)$. As an illustration, we consider a particular type of LWs in $\mathbb{R}^d$
with factorized velocity distribution $F(\bm v) = F_{v}(|\bm v|)\cdot F_{d}(\bm v/|\bm v|)$. In this product PDF first multiplier controls the absolute value of the velocity 
[the simplest  choices is $F_{v}(|\bm v|) = \delta(|\bm v| - v_0)$] while second multiplier 
is governs  the direction statistics of steps. 
A PDF  $F_{d}(\bm v/|\bm v|)$ is a subject of directional statistics \cite{direction} and can be specified with
a probability distribution on the surface of the $(d-1)$ dimensional unit sphere in $\mathbb{R}^d$. For example, in $\mathbb{R}^3$ the continuous transition from the isotropic model to 
the  $XYZ$ LW can be realized with six von Mises-Fisher distributions \cite{direction} (centered at the points where the axes pierce the unit sphere) by tuning
the concentration parameter of the distributions from zero to infinity. We demonstrate in the following sections
that different statistics enter the PDF $P(\bm x,t)$ through the moments $\langle (\bm k\cdot\bm v)^{2n}\rangle$. In particular, for the $XYZ..$ model we have   
\begin{eqnarray}&&\!\!\!\!\!\!\!\!\!\!\!\!\!\!\!
\langle (\bm k\cdot\bm v)^{2n}\rangle=\frac{\sum_{i=1}^d k_i^{2n}}{d}\left(v_0\right)^{2n}. \label{xy}
\end{eqnarray}
In \cite{recent} we discuss physical models belonging to different classes of symmetry, e. g. the Lorentz gas with infinite horizon belongs to the $XYZ..$ class.

The remaining PDF that defines the walk process is $\psi(\tau)$. Below we consider $\psi(\tau)$ that has the tail
\begin{eqnarray}
&& \psi(\tau)\sim \frac{A}{\Gamma(-\alpha)}\tau^{-1-\alpha},\ \ 1<\alpha<2,\label{alph}
\end{eqnarray} 
where $A>0$ and $\Gamma(x)$ is the gamma function (observe that $\Gamma(-\alpha)>0$ when $1<\alpha<2$). The factor $\Gamma(-\alpha)$ 
is introduced in order to make the Laplace transform $\psi(u)$ of $\psi(\tau)$
\begin{eqnarray}
&& \psi(u)=\int_0^{\infty}\exp[-u \tau]\psi(\tau)d\tau,
\end{eqnarray}
to have small $u$ behavior [that is determined by the tail of $\psi(\tau)$],
\begin{eqnarray}&& \psi(u)=1-\langle \tau\rangle u+A u^{\alpha}+\ldots,\label{smlas}
\end{eqnarray}  
where $\langle \tau\rangle=\int_0^{\infty}t\psi(t)dt$ is the average waiting time and dots stand for higher-order terms. 

We use $\bm k$  and $u$ to denote coordinates in Fourier and Laplace space, respectively. By explicitly providing the argument of a function, we will distinguish between the normal or transformed space, for example $\psi(\tau) \rightarrow \psi(u)$ and $g(\bm x) \rightarrow g(\bm k)$.

The lower limit of $\tau$ for which Eq.~(\ref{alph}) holds depends on the considered model. For instance the inverse gamma PDF,
\begin{eqnarray}&&
\psi(\tau)=\frac{2\tau^{-5/2}}{\sqrt{\pi}}\exp\left[-\frac{1}{\tau}\right],
\end{eqnarray}
the tail described by Eq.~(\ref{alph}) holds at $\tau\gg 1$ with $\alpha=3/2$ and $A=8/3$. The corresponding Laplace pair obeys,
\begin{eqnarray}&&\!\!\!\!\!\!\!\!\!
\psi(u)=\left[1+2\sqrt{u}\right]\exp\left[-2\sqrt{u}\right]\sim 1-2u+\frac{8u^{3/2}}{3},
\end{eqnarray}
that reproduces Eq.~(\ref{smlas}) where we use
\begin{eqnarray}&&\!\!\!\!\!\!\!\!\!
\langle \tau\rangle=\int_0^{\infty}\frac{2\tau^{-3/2}d\tau}{\sqrt{\pi}}\exp\left[-\frac{1}{\tau}\right]=\frac{2\Gamma(1/2)}{\sqrt{\pi}}=2.
\end{eqnarray}

We introduce the key instrument of our analysis, the Montroll-Weiss equation. It provides with the Laplace transform 
\begin{eqnarray}
&& P(\bm k, u)=\int_0^{\infty}\exp[-ut] P(\bm k, t) dt,
\end{eqnarray}
of the characteristic function of the position $\bm x(t)$ of the random walker at time $t$,
\begin{eqnarray}&&\!\!\!\!\!\!\!\!\!\!\!\!\! P(\bm k, t)=\langle \exp[i\bm k\cdot\bm x(t)]\rangle=\int \exp[i\bm k\cdot\bm x] P(\bm x, t)d\bm x,
\end{eqnarray}
in terms of averages over statistics of $\bm v$ and $\tau$. We have
\begin{eqnarray}
&&\!\!\!\!\!\!\!\!\!\!\!\!\! P(\bm k, u)=
 \left\langle\frac{1-\psi(u-i\bm k\cdot\bm v)}{u-i\bm k\cdot\bm v}\right\rangle\frac{1}{1-\left\langle \psi(u-i\bm k\cdot\bm v)\right\rangle},\label{basic} \end{eqnarray} 
see Appendix \ref{mw}. Here the angular brackets denote the averaging over the  PDF $F(\bm v)$. The technical problem is to invert this formula in the limit of large time.

\section{Isotropic model}\label{is1d}

In this Section we consider isotropic statistics of velocity where $F(\bm v)$ depends on $|\bm v|$ only. The magnitude of velocity is a random variable drawn from the PDF $S_{d-1} v^{d-1}F(v)$ where $S_{d-1}=2\pi^{d/2}/\Gamma(d/2)$ is the area of unit sphere in $d$ dimensions ($2\pi$ in $d=2$ and $4\pi$ in $d=3$). We show that the PDF $P(\bm x,t)$ of a LW in $\mathbb{R}^d$ can be derived from one-dimensional distribution. 
Thus the well-developed theory of one-dimensional LWs can be used for describing $d$-dimensional isotropic LWs.

We start with observation that for isotropic statistics 
of $\bm v$, the average of an arbitrary function $h$ of $\bm k\cdot\bm v$ depends only on $|\bm k|$; 
thus $\langle h(\bm k\cdot\bm v)\rangle$ can be obtained by taking  $\bm k=k{\hat x}$ (where ${\hat x}$ is unit vector in $x-$direction),
\begin{eqnarray}
&& \langle h(\bm k\cdot\bm v)\rangle=\langle h(k v_x)\rangle=\int_{-\infty}^{\infty} h(k v_x) F(v_x)dv_x,\label{rd}
\end{eqnarray}
where  $F(v_x)$ is the PDF of $x-$component of the velocity. It can be written in terms of the PDF $F(\bm v)= F(v) = F_{d}(\bm v/|\bm v|)$ which obeys the normalization, 
\begin{eqnarray}
&& \int F(\bm v)d\bm v=S_{d-1}\int_0^{\infty} v^{d-1}F(v)dv=1.\label{nrm}
\end{eqnarray}
For $d>2$ we have, 
\begin{eqnarray} \nonumber &&\!\!\!\!\!\!\!\!\! 
F(v_x)=\frac{\int \delta(v_x-v'_x)F(v')d\bm v'}{\int F(v')d\bm v'}\\ \nonumber   &&\!\!\!\!\!\!\!\!\!=\frac{\int_{|v_x|}^{\infty} 
(v')^{d-1}F(v')dv' \int_0^{\pi}\delta(v_x-v'\cos\theta) \sin^{d-2}\theta d\theta}{\int_0^{\infty} (v')^{d-1}F(v')dv' \int_0^{\pi}\sin^{d-2}\theta d\theta}\\ \nonumber  
&&\!\!\!\!\!\!\!\!\!=\frac{\int_{|v_x|}^{\infty} (v')^{d-1}F(v')dv' \int_{-1}^1\delta(v_x-v'x) (1-x^2)^{(d-3)/2} dx}{S_{d-1}^{-1}\int_{-1}^1 (1-x^2)^{(d-3)/2} dx}\nonumber\\&&\!\!\!\!\!\!\!\!\!=
\frac{2\pi^{(d-1)/2}}{\Gamma[(d-1)/2]}\int_{|v_x|}^{\infty} v^{d-2} \left(1-\frac{v_x^2}{v^2}\right)^{(d-3)/2}F(v)dv.\label{xo}
\end{eqnarray}
where we used $\int_{-1}^1 (1-x^2)^{(d-3)/2} dx=\sqrt{\pi}\Gamma[(d-1)/2]/\Gamma(d/2)$ and Eq.~(\ref{nrm}). 
In the case of two dimensions $\theta$ varies between $0$ and $2\pi$ not $\pi$ but the calculation still holds. 
Thus Eq.~(\ref{xo}) provides the distribution of $x-$component of velocity in arbitrary space dimension $d>1$.

Equation~(\ref{xo}) can be simplified further in the case of uniform model with velocity $v_0$ where $F(v)=v_0^{1-d}S_{d-1}^{-1}\delta(v-v_0)$. Integration in Eq.~\eqref{xo} gives 
\begin{eqnarray}&&\!\!\!\!\!\!\!\!\!
F(v_x)=PS_{d/2-2, v_0}(v_x),\nonumber
\end{eqnarray}
where $PS_{d/2-2, v_0}(v)$ is the (normalized) power semicircle PDF with range $v_0$ and shape parameter $d/2-2$ that vanishes when $|v|>v_0$ and for $|v|<v_0$ is given by \cite{psd}
\begin{eqnarray}&&\!\!\!\!\!\!\!\!\!
PS_{d/2-2, v_0}(v)=\frac{\Gamma(d/2)}{\sqrt{\pi}v_0\Gamma[(d-1)/2]} \left(1-\frac{v^2}{v_0^2}\right)^{(d-3)/2}. \label{psd}
\end{eqnarray}
The moments of this distribution read 
\begin{eqnarray}&&\!\!\!\!\!\!\!\!\!\!\!\!\!\!\!\!
\langle |v_x|^{\gamma}\rangle=\frac{v_0^{\gamma}\Gamma[(\gamma+1)/2]\Gamma(d/2)}{\sqrt{\pi}\Gamma[(d+\gamma)/2]}. \label{moments}
\end{eqnarray} 
Note that the ratio of gamma functions  can be rewritten as a product if $d$ is an odd number. 
Finally, from this PDF we can derive the PDF and the moments for arbitrary $F(v)$. For the PDF we find from Eqs.~(\ref{xo}) and (\ref{psd}), 
\begin{eqnarray}&&\!\!\!\!\!\!\!\!\!
F(v_x)=\frac{2\pi^{d/2}}{\Gamma(d/2)}\int_{|v_x|}^{\infty} v^{d-1} PS_{d/2-2, v_x}(v)F(v) dv,\label{fp}
\end{eqnarray}
that can also be seen directly from the definition. For the moments, interchanging the order of integrations, we obtain from Eq.~(\ref{xo}) the identity
\begin{eqnarray}&&\!\!\!\!\!\!\!\!\!\!\!\!\!\!\!\!
\langle |v_x|^{\gamma}\rangle=\int_0^{\infty} v_0^{d-1}S_{d-1} F(v_0)dv_0 \left[\frac{2\pi^{(d-1)/2}}{\Gamma[(d-1)/2]}\right.\nonumber\\&&\!\!\!\!\!\!\!\!\!\!\!\!\!\!\!\!\left.\int\!\! |v_x|^{\gamma}dv_x  \int_{|v_x|}^{\infty} v^{d-2} \!\!\left(1-\frac{v_x^2}{v^2}\right)^{(d-3)/2}\frac{\delta(v-v_0)}{v_0^{d-1}S_{d-1}}dv\right].
\end{eqnarray} 
By noting that the term in brackets is the corresponding moment of $PS_{d/2-2, v_0}$, we find 
\begin{eqnarray}&&\!\!\!\!\!\!\!\!\!\!\!\!\!\!\!\!\langle |v_x|^{\gamma}\rangle\!=\!\frac{2\pi^{(d-1)/2}\Gamma[(\gamma+1)/2]}{\Gamma[(d+\gamma)/2]}\int_0^{\infty} \!\!v^{d+\gamma-1}F(v)dv .\label{xo1}\end{eqnarray} 
For the Gaussian distribution $F(v)=(2\pi {\tilde v}_0^2)^{-d/2}\exp\left[-v^2/2{\tilde v}_0^2\right]$ it gives 
\begin{eqnarray}&&\!\!\!\!\!\!\!\!\!\!\!\!\!\!\!\!\langle |v_x|^{\gamma}\rangle=\frac{2^{\gamma/2}\Gamma[(\gamma+1)/2]{\tilde v}_0^{\gamma}}{\sqrt{\pi}},
\end{eqnarray} 
that reproduces $\langle v_x^2\rangle={\tilde v}_0^2$ when $\gamma=2$. We observe that $\langle |v|^{\gamma}\rangle=S_{d-1}\int_0^{\infty} v^{d+\gamma-1}F(v)dv$ 
so that Eq.~(\ref{xo1}) implies the identity 
\begin{eqnarray}&&\!\!\!\!\!\!\!\!\!\!\!\!\!\!\!\!\langle |v_x|^{\gamma}\rangle=\frac{\Gamma[(\gamma+1)/2]\Gamma(d/2)}{\Gamma[(d+\gamma)/2]\sqrt{\pi}}\langle |v|^{\gamma}\rangle.\label{sp}
\end{eqnarray} 
This formula is a consequence of the isotropy of the process and for any random vector $\bm x$ whose PDF depends on $|\bm x|$ only we have, 
\begin{eqnarray}&&\!\!\!\!\!\!\!\!\!\!\!\!\!\!\!\!\langle |x_s|^{\gamma}\rangle=\frac{\Gamma[(\gamma+1)/2]\Gamma(d/2)}{\Gamma[(d+\gamma)/2]\sqrt{\pi}}\langle |x|^{\gamma}\rangle,\label{sp1}
\end{eqnarray} 
where $x_s,~s \in \{1,2,...,d\}$, is one of the Cartesian coordinates of $\bm x$. 

\begin{figure*}[t]
 \includegraphics[width=0.9\textwidth]{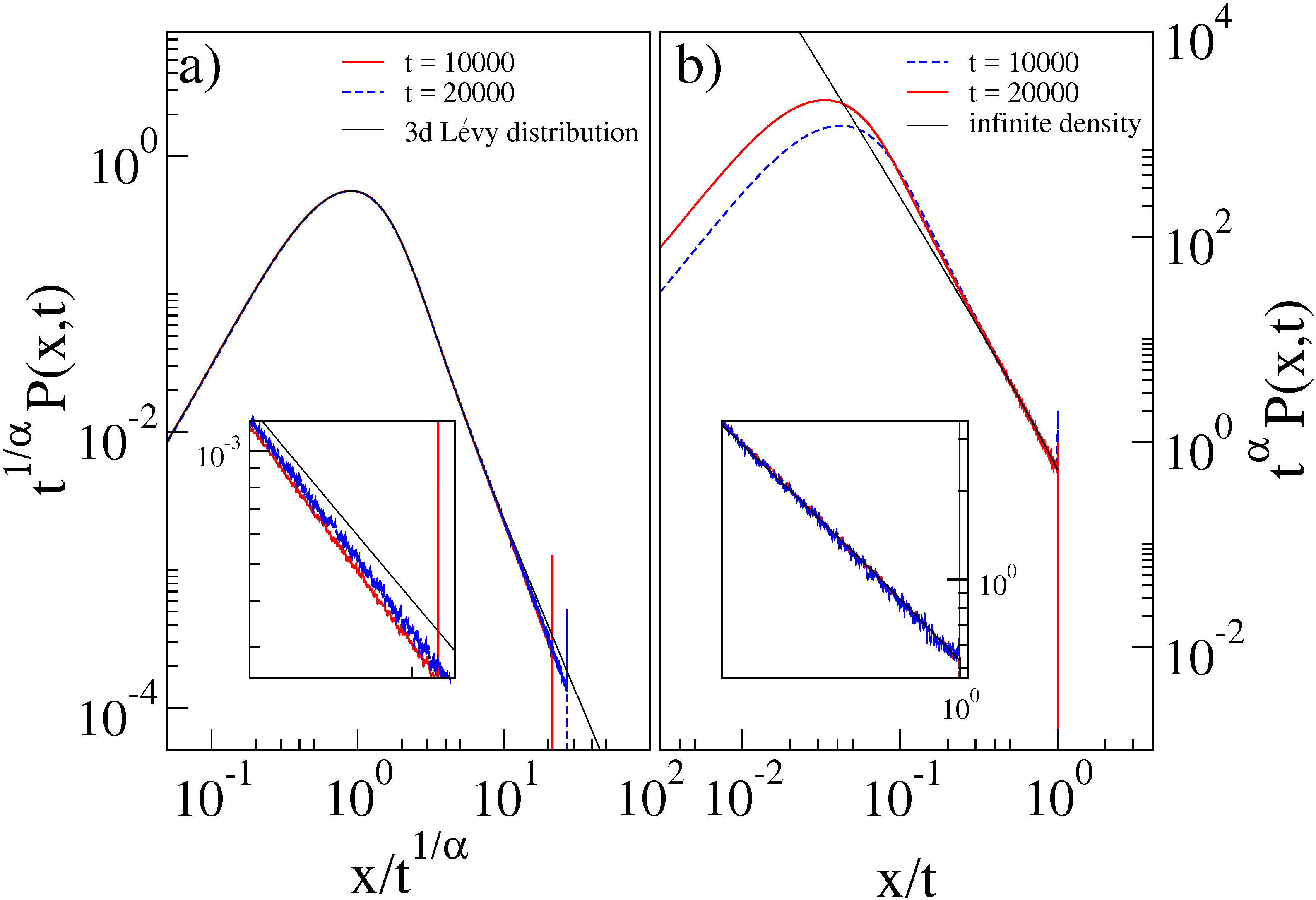}
\caption{\textbf{Scaling limits for a three-dimensional uniform  L\'{e}vy walk.} a) L\'{e}vy scaling of the bulk of numerically sampled PDFs 
$P(x, t)=\left\langle \delta\left(|\bm x(t)|-x\right)\right\rangle$. Thin black line corresponds to the PDFs
$P(x,t)=\left\langle \delta\left(|\bm x(t)|-x\right)\right\rangle_{L_3}$ obtain by averaging over the three-dimensional L\'{e}vy 
distribution, Eq.~(\ref{levy3}); b) Ballistic scaling of the tails of the PDFs. Thin black line is Eq.~(\ref{infisotr}). Both PDFs were sampled over $10^{12}$ realizations.
The parameters are $\alpha = 3/2$, $v_0 = 1$.}
\label{Fig2}
\end{figure*}

\subsection{Bulk statistics: $d-$dimensional L\'{e}vy distributions}

We use Eq.~(\ref{rd}) to rewrite Eq.~(\ref{basic}) in the one-dimensional form, 
\begin{eqnarray}&&\!\!\!\!\!\!\!\!\!\!\!\!\!\! P(k, u)\!=\!\left\langle\frac{1-\psi(u-i k v_x)}{u-ikv_x}\right\rangle_{v_x}\!\!\frac{1}{1-\left\langle \psi(u-ik v_x)\right\rangle_{v_x}},
\end{eqnarray}
where the averaging is taken over the distribution of $v_x$ given  by Eqs.~(\ref{xo}), (\ref{fp}). 
Thus we can directly use the results for $P(k, t)$ in the one-dimensional case. 
The difference from the one-dimensional case is in how the real space PDF is reproduced from $P(k, t)$: here the formula for the inverse Fourier transform of radially symmetric function in $d$ dimensions has to be used. We find from \cite{BarkaiPhysRev} that the bulk of the PDF is described with, 
\begin{eqnarray}&&\!\!\!\!\!\!\!\!\!\!\!\!\!\! P_{cen}(k, t)\sim \exp\left[-K_{\alpha}t|k|^{\alpha}\right],\\&&\!\!\!\!\!\!\!\!\!\!\!\!\!\!
K_{\alpha}=\frac{A}{\langle \tau\rangle} \left\langle |v_x|^{\alpha}\right\rangle\left|\cos\left(\frac{\pi \alpha}{2}\right)\right|,\label{cf}
\end{eqnarray}
where $\left\langle |v_x|^{\alpha}\right\rangle$ is given by Eq.~(\ref{xo1}) with $\gamma=\alpha$. Here the subscript in $P_{cen}$ was introduced in \cite{BarkaiPhysRev}. It stands for the centre (or bulk) part of the PDF. Briefly, to obtain this result we expand $P(k, u)$ using the scaling assumption that $k^{\alpha}$ is of the order $u$ when both are small. Later, we will derive these results as a special case of the more general non-isotropic model (see eq. (\ref{Four}) below).

We can write $K_{\alpha}$ in terms of $\langle |v|^{\alpha}\rangle$ using Eq.~(\ref{sp}), 
\begin{eqnarray}&&\!\!\!\!\!\!\!\!\!\!\!\!\!\!
K_{\alpha}=\frac{A\Gamma[(\alpha+1)/2]\Gamma(d/2)}{\langle \tau\rangle\Gamma[(d+\alpha)/2]\sqrt{\pi}}\langle |v|^{\alpha}\rangle\left|\cos\left(\frac{\pi \alpha}{2}\right)\right|. \label{difc}
\end{eqnarray}
This coefficient reduces to the diffusion coefficient of one-dimensional walk found in \cite{BarkaiPhysRev} setting $d=1$. In the case where $v$ is a conserved constant $v_0$ (modelling conservation of energy), we find  
\begin{eqnarray}&&\!\!\!\!\!\!\!\!\!\!\!\!\!\!
K_{\alpha}=\frac{A\Gamma[(\alpha+1)/2]\Gamma(d/2)}{\langle \tau\rangle\Gamma[(d+\alpha)/2]\sqrt{\pi}}v_0^{\alpha}\left|\cos\left(\frac{\pi \alpha}{2}\right)\right|. 
\end{eqnarray}
Using the inverse Fourier transform we find for the PDF's bulk, 
\begin{eqnarray}&&
P_{cen}(\bm x, t)\sim \frac{1}{(K_{\alpha}t)^{d/\alpha}}L_d\left(\frac{x}{(K_{\alpha}t)^{1/\alpha}}\right),\\&&
L_d(x)=\int \exp[i\bm k\cdot\bm x-k^{\alpha}]\frac{d\bm k}{(2\pi)^d}.\label{symbulk}
\end{eqnarray}
These formulas provide generalized CLT for $d-$dimensional isotropic LWs.  Below these will be generalized to the case of arbitrary (not necessarily isotropic) statistics of velocity. We provide the form of $L_d(x)$ in the cases of physical interest $d=2$ and $d=3$. In the two-dimensional case we have,  
\begin{eqnarray}&&\!\!\!\!\!\!\!\!\! 
L_2(x)=\int_0^{\infty} kJ_0(kx)\exp\left[-k^{\alpha}\right]\frac{dk}{2\pi},
\end{eqnarray}
that can be called two-dimensional isotropic L\'{e}vy density. Here $J_0(z)$ is the Bessel function of the first kind. In three dimensions we find ($L_3(x)$ is normalized $\int L_3(x)d\bm x=1$),
\begin{eqnarray}&&\!\!\!\!\!\!\!\!\!\!\!\!\!\!\!\!\!\!  
L_3(x)\!=\!-\frac{1}{x}(\partial_x)\int_0^{\infty}\!\!\frac{dk}{2\pi^2} \cos(kx)\exp[-k^{\alpha}]\!=\!\frac{L'\left(x\right)}{2\pi x},  \label{levy3}
\end{eqnarray}
where $L(x)=L_1(x)$ is the standard  L\'{e}vy distribution,
\begin{eqnarray}&&\!\!\!\!\!\!\!\!\!\!\!\!\!\!
L(x)=\int \exp\left(ik x-|k|^{\alpha}\right)\frac{dk}{2\pi}. \label{lw1}
\end{eqnarray}
Thus in three dimensions the isotropic L\'{e}vy density can be obtained from the one-dimensional one by differentiation (this is improved version of the old result of \cite{UchZol} valid for arbitrary velocity distribution). This is true for any odd-dimensional case. It can be demonstrated that in the case of even dimension $L_{2n}(x)$ can be obtained from $L(x)$ using derivative operator of half integer order, see Appendix \ref{fraction} and cf. \cite{Marcin}. 

We find from Eq.~(\ref{levy3}) that $L_3(x)\propto |x|^{-\alpha-3}$ at large argument where we use the well-known behavior $L(x)\sim |x|^{-\alpha-1}$, see e. g. \cite{godreche}. (Similarly it will be demonstrated below that $L_d(x)\sim |x|^{-\alpha-d}$.) This tail must fail at larger arguments because it would give divergent dispersion $\langle x^2(t)\rangle\propto \int x^2 L_d(x)d\bm x=\infty$, which is wrong provided that the moments of $F({\bf v})$ are finite, an assumption we use all along this paper. The PDF must necessarily decay fast at $x>v_t t$ where $v_t$ is the typical value of velocity. Thus the L\'{e}vy density does not provide valid description of the tail of the PDF that determines dispersion. This necessitates the study of the tail of the distribution performed below.

\subsection{Infinite density}
The description of the tail of the PDF is performed using the reduction to one-dimensional case where the problem was solved in \cite{BarkaiPhysRev}. This solution is based on asymptotic resummation of the series for the characteristic function. 

We observe that for isotropic statistics of velocity $P(k, t)$ obeys,
\begin{eqnarray}&&\!\!\!\!\!\!\!\!\!\!\!\!\!\! P(k, t)=\left\langle \exp\left[-i\bm k\cdot\bm x\right ]\right\rangle=1+\sum_{n=1}^{\infty} \frac{(-1)^n k^{2n}\left\langle x_1^{2n}\right\rangle}{(2n)!},\label{ch}
\end{eqnarray}
where we used that odd moments of $\bm k\cdot \bm x$ vanish and that isotropy implies that $\left\langle \left(\bm k\cdot\bm x\right)^{2n}\right\rangle$ is independent of direction of $\bm k$ so it can be obtained setting $\bm k=k{\hat x}$ giving $\left\langle \left(\bm k\cdot\bm x\right)^{2n}\right\rangle=k^{2n}\langle x_1^{2n}\rangle$, cf. with similar consideration for velocity. 

Though $\left\langle x_1^{2n}\right\rangle$ cannot be found completely at all times, it was discovered in \cite{BarkaiPhysRev} that this can be done asymptotically in the limit of large times. The proper adaptation of the result tells that using small $k$ and $u$ expansion of the quasi-one-dimensional Montroll-Weiss equation (\ref{basic}) when keeping the ratio $k/u$ fixed we find,
\begin{eqnarray}&&\!\!\!\!\!\!\!\!\!\!\!\!\!\! P(k, t)\!\sim \!1\!+\!\frac{A}{\langle \tau\rangle}\!\sum_{n=1}^{\infty} \!\frac{\Gamma(2n\!-\!\alpha)(-1)^n  t^{2n\!+\!1\!-\!\alpha}\!\left\langle v_x^{2n}\right\rangle\! k^{2n}}{(2n\!-\!1)!|\Gamma(1\!-\!\alpha)|\Gamma(2n\!+\!2\!-\!\alpha)}.\label{istr0}\end{eqnarray}
This result was obtained in the limit of long times asymptotically that is the $n-$th term in the series is valid provided time is large. How large this time is depends on $n$: the higher $n$ is, the larger times are needed for the validity of the asymptotic form. Thus at however large but finite $t$ the terms of the series fail starting from some large but finite $n$.  Thus, in contrast, to Eq.~(\ref{ch}) the resummation of the series given by Eq.~(\ref{istr0}) does not have to lead to $P(x, t)$ because there is no time for which all terms in the series of Eq.~(\ref{istr0}) are valid.

It is the finding of \cite{BarkaiPhysRev} that resummation of the series in Eq.~(\ref{istr0}) still produces function that has physical meaning. That function called the infinite density gives valid description of the tail of $P(x, t)$ but fails in the bulk. This is because the bulk corresponds to small $x$ and large $k\propto 1/x$. For very small $x$ very large $k$ are relevant implying that terms with very large $n$ become relevant for the sum. However these terms are not valid at finite $t$ leaving the small $x$ inaccessible for the sum in Eq.~(\ref{istr0}).

Since statistics of both $\bm x$ and $\bm v$ are isotropic then,
\begin{eqnarray}&&\!\!\!\!\!\!\!\!\!\!\!\!\!\! \frac{\langle x_1^{2n}\rangle}{\langle x^{2n}\rangle}=\frac{\langle v_x^{2n}\rangle}{\langle v^{2n}\rangle}, \label{istr}
\end{eqnarray}
see Eq.~(\ref{sp1}). We find comparing the series in Eqs.~(\ref{ch})-(\ref{istr0}),
\begin{eqnarray}&&\!\!\!\!\!\!\!\!\!\!\!\!\!\! 
\left\langle x^{2n}(t)\right\rangle\!=\!\frac{2nA \left\langle v^{2n}\right\rangle}{\langle \tau\rangle|\Gamma(1-\alpha)|(2n\!+\!1\!-\!\alpha)(2n-\!\alpha)}t^{2n+1-\alpha}.\label{moments}
\end{eqnarray}
Remarkably this is independent of dimension and thus coincides with the one-dimensional case 
(isotropy implies that the geometry disappears from $\langle x^{2n}\rangle$ because of Eq.~(\ref{istr})). 
The use of these moments for formal reconstruction of the long-time limit of the 
PDF $P(x, t)=\left\langle \delta\left(|\bm x(t)|-x\right)\right\rangle$ of $|\bm x(t)|$ through $P_A(x, t)$ defined by,
\begin{eqnarray}&&\!\!\!\!\!\!\!\!\!\!\!\!\!\!\!\!
P_A(x, t)\!=\!\!\int\!\! \frac{dk}{2\pi} \exp[-ikx]\left[1\!+\!\sum_{n=1}^{\infty} \frac{(ik)^{2n}\left\langle x^{2n}(t)\right\rangle}{(2n)!}\right],~
\end{eqnarray}
with $\left\langle x^{2n}(t)\right\rangle$ given by Eq.~(\ref{moments}) gives on resummation \cite{BarkaiPhysRev},
\begin{eqnarray}&&\!\!\!\!\!\!\!\!\!\!\!\!\!\! 
P_A(x, t)=\frac{AS_{d-1}}{\langle \tau\rangle |\Gamma(1-\alpha)|t^{\alpha}}\int_{|x|/t}^{\infty} v^{d-1}F(v)dv\nonumber\\&&\!\!\!\!\!\!\!\!\!\!\!\!\!\!\times\left[\alpha\frac{|v|^{\alpha}}{|x/t|^{1+\alpha}}
-(\alpha-1)\frac{|v|^{\alpha-1}}{|x/t|^{\alpha}}\right], \label{infisotr}
\end{eqnarray} that holds for $x\neq 0$. The function $P_A(x, t)$ clearly describes the long-time behavior of the moments of $|\bm x(t)|$ via 
\begin{eqnarray}&&\!\!\!\!\!\!\!\!\!\!\!\!\!\! 
\langle x^{2n}(t)\rangle\sim \int_0^{\infty} P_A(x, t) x^{2n}dx,\ \ n\geq 1.
\end{eqnarray}
This function however does not describe the normalization (obtained as $n=0$) 
since $P_A(x, t)\sim x^{-(1+\alpha)}$ for $x\to 0$. Hence $P_A(x, t)$ is not normalizable for which 
reason it is called infinite density. However it does describe the integer order moments $\int_0^{\infty} P(x, t) x^{2n}dx$ where $P(x, t)$ 
is the PDF of the distance to the origin $|\bm x(t)|$. Thus $P(x, t)\sim P_A(x, t)$ is true for integrals with integer powers. 
It can be seen that this function describes the tail of the PDF $P(x, t)$ at $x\sim t$ (that is at large times $P(x, t)\sim P_A(x, t)$ 
holds for large $x\propto t$) while the bulk corresponds to $x\sim t^{1/\alpha}$. This fits that the moments of integer order are 
determined by the tail of $P(x, t)$ as clarified in the coming Sections. Below we derive the infinite density in $d$ dimensions in 
different form proving the existence of finite long-time limit $\lim_{t\to\infty} t^{d-1+\alpha}P(t\bm x, t)$ for arbitrary (anisotropic) 
statistics of velocity. The descriptions of the bulk and the tail of the PDF together with the moments provide a complete description of the $d-$dimensional 
walk with isotropic statistics. 

\section{CLT for anisotropic L\'{e}vy walks}\label{CLT}

We start the analysis of the anisotropic LWs with derivation of the generalized CLT that describes PDF $P(\bm x,t)$ of the walker. We consider in this Section random walks whose single step duration's PDF $\psi(u)$ obeys Eq.~(\ref{smlas}) at small $u$ but we let the range of considered $\alpha$ include $\alpha=2$. That is we consider $\psi(u)$ in Eq.~(\ref{smlas}) with $1<\alpha\leq 2$. In the case of $\alpha=2$ though $\psi(\tau)$ does not obey Eq.~(\ref{alph}) with $\alpha=2$. This is because Eq.~(\ref{smlas}) with $\alpha=2$ describes the case of finite dispersion of $\tau$ given by $\langle \tau^2\rangle=\psi''(u=0)=2A$. In contrast, for $\alpha=2$ Eq.~(\ref{alph}) gives $\tau^{-3}$ tail for which the dispersion is infinite. Thus $\psi(\tau)$ obeying Eq.~(\ref{smlas}) with $\alpha=2$ decays faster than $\tau^{-3}$. 

We demonstrate that for $\psi(u)$ obeying Eq.~(\ref{smlas}) with $1<\alpha\leq 2$ there is finite limit, 
\begin{eqnarray}&&\!\!\!\!\!\!\!\!\!\!\!\!\!\!
\lim_{t\to\infty} t^{d/\alpha}P(t^{1/\alpha}\bm x, t)\nonumber\\&&\!\!\!\!\!\!\!\!\!\!\!\!\!\!=\int \exp\left[i\bm k\cdot\bm x\!-\!\frac{A}{\langle \tau\rangle}\left|\cos\left(\frac{\pi\alpha}{2}\right)\right|\left\langle |\bm k\cdot\bm v|^{\alpha}\right\rangle \right]\frac{d\bm k}{(2\pi)^d}, \label{lw} 
\end{eqnarray}
that holds for arbitrary statistics of $\bm v$ obeying $F(\bm v)=F(-\bm v)$. It is proper to call this result generalized CLT because for $\alpha=2$ it reproduces the central limit theorem,
\begin{eqnarray}&&\!\!\!\!\!\!\!\!\!\!\!\!\!\!
\!\lim_{t\to\infty}\!t^{d/2}P(\bm x\sqrt{t}, t)\!=\!\!\int \!\!\exp\left[i\bm k\cdot\bm x\!-\!\frac{\Gamma_{pl} k_pk_l}{2}\right]\!\frac{d\bm k}{(2\pi)^d},\label{rescl}
\end{eqnarray}
where the RHS defines $g(\bm x)$ in Eq.~(\ref{g}), the covariance matrix $\Gamma$ is defined by, 
\begin{eqnarray}&&\!\!\!\!\!\!\!\!\!\!\!\!\!\!
\Gamma_{pl}=\frac{\langle\tau^2\rangle\langle v_pv_l\rangle}{\langle\tau\rangle},
\end{eqnarray}
and we used $A=\langle\tau^2\rangle/2$. In Eq.~(\ref{rescl}) we use Einstein summation rule over the repeated indices. We observe that the units of $\bm k$ in Eq.~(\ref{rescl}) are $t^{1/2}/l$, units of $\bm x$ are $l/t^{1/2}$ and units of $\Gamma$ are $l^2/t$ so that the argument of the exponent is dimensionless.

The form of the covariance matrix could be seen considering the second moment of displacement $\bm x(t)\!=\!\sum_{k=1}^{N(t)} \bm v_k \tau_k$,
\begin{eqnarray}&&\!\!\!\!\!\!\!\!\!\!\!\!\!\!
\lim_{t\to\infty}\frac{\langle x_p(t)x_l(t)\rangle}{t}\!=\!\lim_{t\to\infty}\frac{\langle N(t)v_pv_l\tau^2\rangle}{t}\!=\!\frac{\langle\tau^2\rangle\langle v_pv_l\rangle}{\langle \tau\rangle},\nonumber
\end{eqnarray} where we used that the law of large numbers implies $\lim_{t\to\infty}\langle N(t)\rangle/t=1/\langle\tau\rangle$.

In the Gaussian $\alpha=2$ case the details of statistics of individual steps of the walk become irrelevant in the long-time limit: they get summarized in $d(d+1)/2$ independent coefficients of the covariance matrix $\Gamma$. The second moments of velocity $\langle v_pv_l\rangle$ determine the long-time statistics of the displacement uniquely. In contrast in $\alpha<2$ case the details of the walk influence the displacement's PDF however large time is via $\left\langle |\bm k\cdot\bm v|^{\alpha}\right\rangle$. This is non-trivial function of direction of $\bm k$ that depends on which directions of motion are more probable in one step. This is a function of continuous variable rather than $\Gamma$ that depends on finite number of discrete indices. Here we assume that isotropy is broken - in the isotropic case the degree of universality of $\alpha<2$ and $\alpha=2$ is the same: $\langle |v|^{\alpha}\rangle$ determines uniquely the long-time behavior given by Eq.~(\ref{difc}). Correspondingly infinite variability of shapes of $P(\bm x, t)$ is possible in contrast with fixed Gaussian shape in $\alpha=2$ case. 

\begin{figure*}[t]
\includegraphics[width=0.9\textwidth]{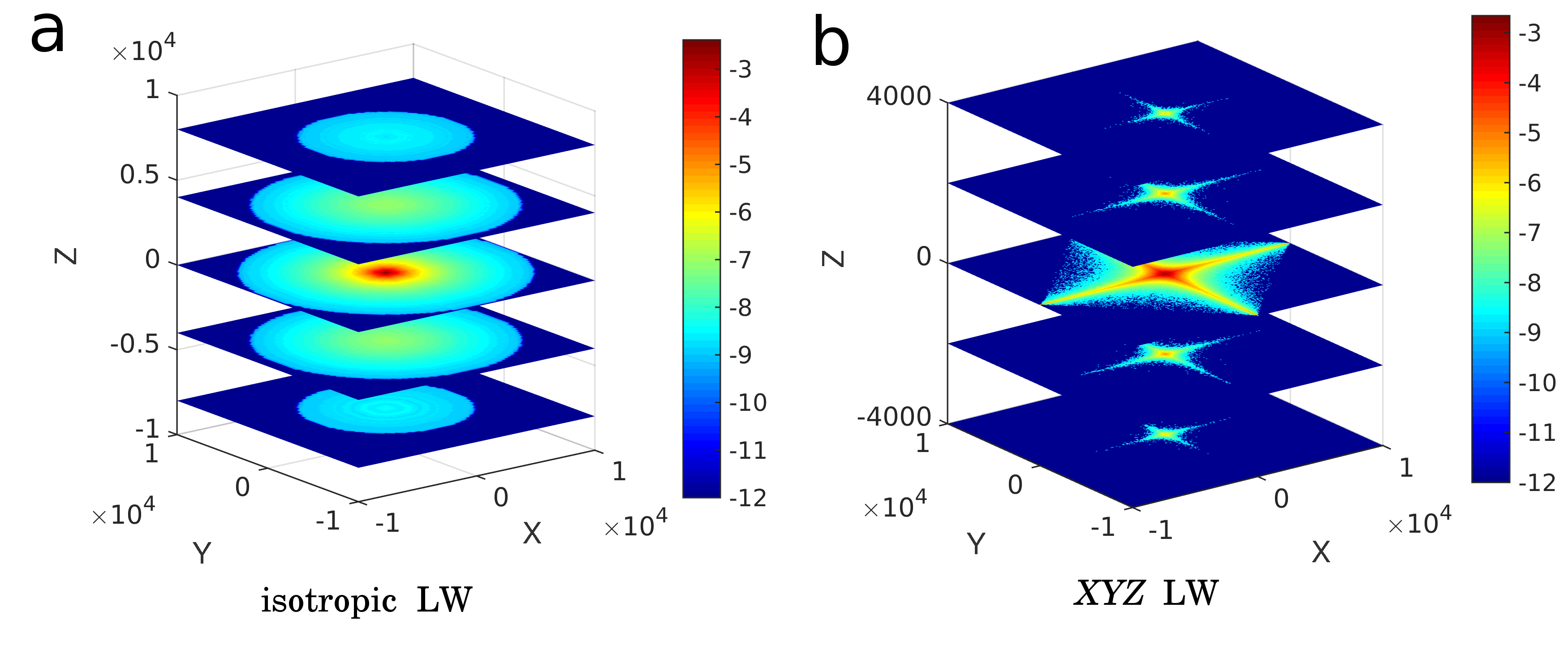}
\caption{\textbf{Probability density functions of Levy walks in three-dimensions.}
The distributions  for
the time $t/\tau_{0} = 10^4$ were obtained by sampling over $10^{11}$ realizations. They are represented by  the set of two-dimensional
'slices' along $z$-axis, $P(x,y,z=const)$,  plotted on a log scale. The other parameters are as in Fig. 1.}
\label{Fig3}
\end{figure*}

We start the derivation of Eq.~(\ref{lw}). The calculation below hold for $1<\alpha\leq 2$. We use 
\begin{eqnarray}&&\!\!\!\!\!\!\!\!\!\!\!\!\!\!
t^{d/\alpha}P(t^{1/\alpha}\bm x, t)=t^{d/\alpha} \int\frac{d\bm k}{(2\pi)^d}\exp\left[it^{1/\alpha}\bm k\cdot\bm x\right]P(\bm k, t)\nonumber\\&&\!\!\!\!\!\!\!\!\!\!\!\!\!\!=\int\frac{d\bm k}{(2\pi)^d}\exp\left[i\bm k\cdot\bm x\right]P(t^{-1/\alpha}\bm k, t).
\end{eqnarray}
Thus we have to prove the existence of the limit, 
\begin{eqnarray}&&\!\!\!\!\!\!\!\!\!\!\!\!\!\!
\lim_{t\to\infty}\! P(t^{-1/\alpha}\bm k, t)\!=\! \!\int \!\frac{du}{2\pi i} \exp[u]\lim_{t\to\infty}\!\frac{1}{t}P\left(\frac{\bm k}{t^{1/\alpha}}, \frac{u}{t}\right).\label{lmt}
\end{eqnarray}
We use that ($\langle \bm k\cdot\bm v\rangle=0$),
\begin{eqnarray}&& \!\!\!\!\!\!\!\!\!\!\!\!\!\!\left\langle\frac{1-\psi(ut^{-1}-i\bm k\cdot\bm vt^{-1/\alpha})}{ut^{-1}-i\bm k\cdot\bm vt^{-1/\alpha}}\right\rangle=\langle \tau\rangle+o(t),\label{lmt1}\\&&\!\!\!\!\!\!\!\!\!\!\!\!\!\!1\!-\!\left\langle \psi\left(\frac{u}{t}\!-\!\frac{i\bm k\cdot\bm v}{t^{1/\alpha}}\right)\right\rangle\!=\!\frac{\langle \tau\rangle u}{t}\!-\!A\left\langle \left(
\frac{u}{t}\!-\!\frac{i\bm k\cdot\bm v}{t^{1/\alpha}}\right)^{\alpha}\right\rangle\\&&\!\!\!\!\!\!\!\!\!\!\!\!\!\!+o(t)=\frac{\langle \tau\rangle u+A|\cos(\pi\alpha/2)|\left\langle |\bm k\cdot\bm v|^{\alpha}\right\rangle}{t},\label{lmt2}
\end{eqnarray}
where we used that $F(\bm v)=F(-\bm v)$ implies \cite{BarkaiPhysRev},
\begin{eqnarray}&& \!\!\!\!\!\!\!\!\!\!\!\!\!\!\left\langle \left(-i\bm k\cdot\bm v\right)^{\alpha}\right\rangle=\left\langle |\bm k\cdot\bm v|^{\alpha}\exp\left[-\frac{i\pi \alpha}{2} sign\left(\bm k\cdot\bm v\right)\right]\right\rangle\nonumber\\&&\!\!\!\!\!\!\!\!\!\!\!\!\!\!=\left\langle |\bm k\cdot\bm v|^{\alpha}\right\rangle \cos\left(\frac{\pi\alpha}{2}\right).\label{branch}
\end{eqnarray}
This formula holds in $\alpha=2$ case as well. We find using Eqs.~(\ref{lmt1})-(\ref{lmt2}) in the Montroll-Weiss equation that,
\begin{eqnarray}&&\!\!\!\!\!\!\!\!\!\!\!\!\!\!
\lim_{t\to\infty}\frac{1}{t}P\left(\frac{\bm k}{t^{1/\alpha}}, \frac{u}{t}\right)=\frac{1}{u+{\tilde A}|\cos(\pi\alpha/2)|\left\langle |\bm k\cdot\bm v|^{\alpha}\right\rangle},\nonumber
\end{eqnarray}
where ${\tilde A}=A/\langle\tau\rangle$, cf. one-dimensional case in \cite{BarkaiPhysRev}. We conclude that, 
\begin{eqnarray}&&\!\!\!\!\!\!\!\!\!\!\!\!\!\!
\!\lim_{t\to\infty}\!\! P(t^{-1/\alpha}\bm k, t)\!=\!\! \!\int \!\frac{du}{2\pi i} \frac{\exp[u]}{u+{\tilde A}|\cos(\pi\alpha/2)|\left\langle |\bm k\cdot\bm v|^{\alpha}\right\rangle},
\end{eqnarray}
see Eq.~(\ref{lmt}). We find performing the integration, 
\begin{eqnarray}&&\!\!\!\!\!\!\!\!\!\!\!\!\!\! \!\lim_{t\to\infty}\!\! P(t^{-1/\alpha}\bm k, t)\!=\exp\left[-\frac{A}{\langle\tau\rangle}\left|\cos\left(\frac{\pi\alpha}{2}\right)\right|\left\langle |\bm k\cdot\bm v|^{\alpha}\right\rangle\right],
\end{eqnarray}completing the proof of Eq.~(\ref{lw}). We find asymptotically at large times that,
\begin{eqnarray}&&\!\!\!\!\!\!\!\!\!\!\!\!\!\!
P(\bm k, t)\sim \exp\left[-\frac{t A}{\langle\tau\rangle}\left|\cos\left(\frac{\pi\alpha}{2}\right)\right|\left\langle |\bm k\cdot\bm v|^{\alpha}\right\rangle\right].\label{Four}
\end{eqnarray}
This is one of our main results as it provides the generalized CLT for non-isotropic LWs. For isotropic case Eq.~(\ref{Four}) reduces to Eqs.~(\ref{cf}).

We characterize different statistics of velocity with structure function $s({\hat k})$ that depends on the unit vector ${\hat k}=\bm k/k$, 
\begin{eqnarray}&&\!\!\!\!\!\!\!\!\!\!\!\!\!\!
s({\hat k})=\frac{\Gamma[(d+\alpha)/2]\sqrt{\pi}}{\Gamma[(\alpha+1)/2]\Gamma(d/2)} \frac{\left\langle |{\hat k}\cdot\bm v|^{\alpha}\right\rangle}{\langle |v|^{\alpha}\rangle}. 
\end{eqnarray}
This is defined so that for isotropic statistics $s({\hat k})=1$ (in that case we have $\left\langle |{\hat k}\cdot\bm v|^{\alpha}\right\rangle=\left\langle |v_x|^{\alpha}\right\rangle$ where $\left\langle |v_x|^{\alpha}\right\rangle$ is determined by Eq.~(\ref{sp})). We have with this definition, 
\begin{eqnarray}&&\!\!\!\!\!\!\!\!\!\!\!\!\!\!
\lim_{t\to\infty} t^{d/\alpha}P(t^{1/\alpha}\bm x, t)\!=\!\!\int \!\!\exp\left[i\bm k\!\cdot\!\bm x\!-\!K_{\alpha} k^{\alpha} s({\hat k})\right]\!\frac{d\bm k}{(2\pi)^d}, \nonumber \\&&\!\!\!\!\!\!\!\!\!\!\!\!\!\!
P(\bm k, t)\sim \exp\left[-K_{\alpha} t k^{\alpha} s({\hat k})\right],\label{lw10}
\end{eqnarray}
where we use $K_{\alpha}$ defined in Eq.~(\ref{difc}) (in anisotropic case $K_{\alpha}$ does not have direct interpretation of diffusion coefficient so this is to be taken as mathematical definition). The PDF of the displacement obeys,
\begin{eqnarray}&&\!\!\!\!\!\!\!\!\!\!\!\!\!\!
P(\bm x, t)\sim \frac{1}{(K_{\alpha}t)^{d/\alpha}}{\hat L}_d\left(\frac{\bm x}{(K_{\alpha}t)^{1/\alpha}}\right),\label{distr}\\&&\!\!\!\!\!\!\!\!\!\!\!\!\!\!
{\hat L}_d(\bm x)\!=\!\int\!\! \exp\left[i\bm k\cdot\bm x\!-k^{\alpha} s({\hat k})\right]\frac{d\bm k}{(2\pi)^d}. \label{distrl}
\end{eqnarray}
For isotropic model there is no modulation and ${\hat L}_d(\bm x)$ is the universal function $L_d(x)$ introduced previously, see Eq.~(\ref{symbulk}). Thus different isotropic statistics produces the same long-time PDF in the bulk that differ only by the value of $K_{\alpha}$. For $XYZ...$ model $\left\langle |\bm k\cdot\bm v|^{\alpha}\right\rangle=v_0^{\alpha}\sum_{i=1}^d |k_i|^{\alpha}/d$ we find that the distribution factorizes in the product of one-dimensional distributions. Thus in this case the bulk of the PDF coincides with that of independent walks along different axes. In these cases the functional shape is universal. 

The chief feature introduced by the passage from one dimension to the higher-dimensional case is that quite arbitrary angular structure of the distribution becomes possible. The structure function $s({\hat k})$ describes positive angular modulation in $\bm k-$space that changes correspondingly the functional form in real space, see Eq.~(\ref{distrl}). This function does not seem to obey strong constraints that would strongly limit the possible forms of ${\hat L}_d(\bm x)$. We stress this fact calling distribution ${\hat L}_d(\bm x)$ 'anisotropic $d-$dimensional L\'{e}vy distribution' in contrast with $d-$dimensional isotropic L\'{e}vy distributions introduced previously \cite{UchZol} in the context of L\'evy flights that have universal shape. 

Considering marginal distributions of components of $\bm x$ one reduces the problem to one dimension restoring universality of the distribution. Integrating Eq.~(\ref{distr}),
\begin{eqnarray}&&\!\!\!\!\!\!\!\!\!\!\!\!\!\!
P(x_i, t)\!=\!\!\int\!\! P(\bm x, t)\prod_{k\neq i} dx_k\!\sim \!\frac{1}{(K_it)^{1/\alpha}}L\left(\frac{x_i}{(K_it)^{1/\alpha}}\right),
\end{eqnarray}
where $L(x)$ is defined in Eq.~(\ref{lw1}) and $K_i$ is "diffusion coefficient in $i-$th direction", 
\begin{eqnarray}&&\!\!\!\!\!\!\!\!\!\!\!\!\!\!
K_i=\!\frac{A}{\langle\tau\rangle}\left|\cos\left(\frac{\pi\alpha}{2}\right)\right|\left\langle |v_i|^{\alpha}\right\rangle.
\end{eqnarray}
Thus marginal PDFs are given by the standard one-dimensional symmetric L\'{e}vy stable law. In contrast the PDF $P(x, t)=\int \delta(|\bm x(t )|-x)P(\bm x, t)d\bm x$ of the distance $x(t)=|\bm x(t)|$ from the origin is not universal. We find integrating Eq.~(\ref{distr}) that,
\begin{eqnarray}&&\!\!\!\!\!\!\!\!\!\!\!\!\!\!
P(x, t)\sim \frac{1}{t^{1/\alpha}}P_0\left(\frac{x}{t^{1/\alpha}}\right),\label{rm}\\&&\!\!\!\!\!\!\!\!\!\!\!\!\!
P_0(x)\!=\! \!\!\int\!\! \frac{x^{d/2} J_{d/2-1}(kx) d\bm k} {(2\pi)^{d/2}k^{d/2-1}} \!\exp\left[\!-K_{\alpha} k^{\alpha}s({\hat k}) \right]\! 
,\label{pi0}
\end{eqnarray}
(see further details for notation choices in the next Section) where we used 
\begin{eqnarray}&&\!\!\!\!\!\!\!\!\!\!\!\!\!\!
\int \exp\left[i\bm k\cdot\bm x\right] \delta(|\bm x(t)|-x) d\bm x= \frac{(2\pi x)^{d/2}J_{d/2-1}(kx)}{k^{d/2-1}},\nonumber
\end{eqnarray} where $J_{\nu}(z)$ is the Bessel function of the first kind of order $\nu$. We observe that integration over angles implied by the definition of the PDF of the distance from the origin does not bring universal
form of the PDF (that would then coincide with the PDF for isotropic statistics). The structure function is present in Eq.~(\ref{pi0}) and can produce quite different forms of $P(x, t)$. We have using Taylor series for the Bessel function in Eq.~(\ref{pi0}),
\begin{eqnarray}&&\!\!\!\!\!\!\!\!\!\!\!\!\!\!
P_0(x)= x^{d-1}\sum_{n=0}^{\infty}\frac{(-1)^n c_n}{n!\Gamma(d/2+n)2^{d/2-1}}\left(\frac{x}{2}\right)^{2n},\\&&\!\!\!\!\!\!\!\!\!\!\!\!\!c_n= 
\int\!\! \frac{d\bm k} {(2\pi)^{d/2}} k^{2n}\exp\left[-K_{\alpha}k^{\alpha}s({\hat k}) \right],\nonumber
\end{eqnarray}
where $x^{d-1}$ factor describes the contribution of the surface of the sphere of radius $x$. In the isotropic case we have,  
\begin{eqnarray}&&\!\!\!\!\!\!\!\!\!\!\!\!\!\!
c_n\!=\!\int\!\! \frac{d\bm k} {(2\pi)^{d/2}} k^{2n}\exp\left[-K_{\alpha} k^{\alpha} \right]\!=\!\frac{S_{d-1}\Gamma[(2n\!+\!d)/\alpha]}{(2\pi)^{d/2}\alpha K_{\alpha}^{(2n\!+\!d)/\alpha}}\nonumber.
\end{eqnarray}
We find using that in isotropic case $P_0(\bm x)=P_0(x)/[x^{d-1}S_{d-1}]$ the Taylor series for $P_{cen}(\bm x, t)$ defined in Eq.~(\ref{symbulk}), 
\begin{eqnarray}&&
L_d(x)=\sum_{n=0}^{\infty}\frac{(-1)^n \Gamma[(2n\!+\!d)/\alpha]}{n!\Gamma(n+d/2)2^{2n+d-1}\pi^{d/2}\alpha}x^{2n},\label{lvtaylor}
\end{eqnarray}
The factor of $\Gamma(n+d/2)$ can be simplified in cases of odd and even dimensions using $\Gamma(n+1)=n!$ and $\Gamma(n+1/2)=2^{-n}\sqrt{\pi}(2n-1)!!$. In the case of $d=1$ the series reproduces the one-dimensional formula \cite{godreche}. 

As mentioned, for arbitrary statistics of velocity there seems to be no constraint on the Taylor coefficients $c_n$ that would determine uniquely the functional form of $P_0(x)$. However despite that the small $x$ expansion of $P_0(x)$ is not universal, the large $x$ behavior of $P_0(x)$ is universal. This will be demonstrated in the next Section. 

\section{Universal tail of anisotropic L\'{e}vy distribution and low-order moments}\label{ls}

In this Section we demonstrate that the PDF of the distance of the walker from the walk's origin has power-law tail with universal (independent of statistics of velocity) exponent. This has the implication that moments of distance from the origin with order smaller than $\alpha$ are determined by the bulk of the PDF but moments of higher order are determined by the PDF's tail (whose description is different from L\'{e}vy distribution and is provided further in the text). Thus dispersion is determined by the tail of the PDF independently of the statistics of velocity.   

We observe from Eq.~(\ref{pi0}) that, 
\begin{eqnarray}&&\!\!\!\!\!\!\!\!\!\!\!\!\!\!
P_0(x)-\delta(\bm x)= x^{d/2}\int\!\! \frac{d\bm k} {(2\pi)^{d/2}}k^{1-d/2}J_{d/2-1}(kx) \nonumber\\&&\!\!\!\!\!\!\!\!\!\!\!\!\!\times \left(\exp\left[-\!\frac{ A}{\langle\tau\rangle}\left|\cos\left(\frac{\pi\alpha}{2}\right)\right|\left\langle |\bm k\cdot\bm v|^{\alpha}\right\rangle \right]-1\right),\nonumber
\end{eqnarray}
where we used the  Fourier transform representation of the $\delta-$function and restored the definitions of the constants.
By using the large argument asymptotic expansion of Bessel function and rescaling integration variable by $x$, we find
\begin{eqnarray}&&\!\!\!\!\!\!\!\!\!\!\!\!\!\!
P_0(x)\!\sim\! \frac{\sqrt{2}}{x\sqrt{\pi}}\int\!\! \frac{d\bm k} {(2\pi)^{d/2}} k^{(1-d)/2}\cos\left(k-\frac{\pi (d-1)}{4}\right)\nonumber\\&&\!\!\!\!\!\!\!\!\!\!\!\!\!\times\left(\exp\left[-\!\frac{ A\left\langle |\bm k\cdot\bm v|^{\alpha}\right\rangle}{x^{\alpha}|\langle\tau\rangle}\left|\cos\left(\frac{\pi\alpha}{2}\right)\right| \right]-1\right)\exp\left[-\epsilon k\right],\label{xp}
\end{eqnarray}
where infinitesimal $\epsilon$ in the last term is the convergence factor introduced for convergence of the large $x$ expansion found by expanding the exponent in brackets. The leading order term is,
\begin{eqnarray}&&\!\!\!\!\!\!\!\!\!\!\!\!\!\!
P_0(x)\!\sim \! -\frac{A\sqrt{2}}{x^{1+\alpha}\sqrt{\pi}\langle\tau\rangle}\left|\cos\left(\frac{\pi\alpha}{2}\right)\right|\int\!\! \frac{ k^{(1-d)/2}d\bm k} {(2\pi)^{d/2}}\nonumber\\&&\!\!\!\!\!\!\!\!\!\!\!\!\!\times \left\langle |\bm k\cdot\bm v|^{\alpha}\right\rangle \cos\left(k-\frac{\pi (d-1)}{4}\right)\exp\left[-\epsilon k\right].\label{tail}
\end{eqnarray}
The angular integral gives 
\begin{eqnarray}&&\!\!\!\!\!\!\!\!\!\!\!\!\!\!
\int d{\hat k} \left\langle |{\hat k}\cdot\bm v|^{\alpha}\right\rangle=\left\langle v^{\alpha} \right\rangle\int d{\hat k} |{\hat k}\cdot{\hat v}|^{\alpha}, \label{ang}
\end{eqnarray}
where we interchanged the orders of averaging and integration and used that the last integral is independent of ${\hat v}$. 
However since it is independent of ${\hat v}$ then it can be obtained taking ${\hat v}$ in $x-$direction which gives, 
\begin{eqnarray}&&\!\!\!\!\!\!\!\!\!\!\!\!\!\!
\int d{\hat k} \left\langle |{\hat k}\cdot\bm v|^{\alpha}\right\rangle=\left\langle v^{\alpha} \right\rangle\int d{\hat k} |{\hat k}_x|^{\alpha},
\end{eqnarray}
We find using this in Eq.~(\ref{tail}), 
\begin{eqnarray}&&\!\!\!\!\!\!\!\!\!\!\!\!\!\!
P_0(x)\!\sim \! -\frac{\sqrt{2}\Gamma[(d+\alpha)/2]K_{\alpha}}{x^{1+\alpha}\Gamma[(\alpha+1)/2]\Gamma(d/2)}\int\!\! \frac{ k^{(1-d)/2}d\bm k} {(2\pi)^{d/2}}\nonumber\\&&\!\!\!\!\!\!\!\!\!\!\!\!\!\times |k_x|^{\alpha}  \cos\left(k-\frac{\pi (d-1)}{4}\right)\exp\left[-\epsilon k\right],
\end{eqnarray}
where we use $K_{\alpha}$ defined in Eq.~(\ref{difc}). It is seen readily from Eq.~(\ref{xp}) that the condition of applicability of large $x$ expansion is $x\gg K_{\alpha}^{1/\alpha}$ (cf. Eq.~(\ref{rm}) for checking the units). The $x-$axis defining $k_x$ in the integrand is arbitrary direction in space. We observe that $|k_x|^{\alpha}$ averaged over the angles can be found using in Eq.~(\ref{sp1}) the isotropic statistics with $\bm x$ switched by $\bm k$:
\begin{eqnarray}&&\!\!\!\!\!\!\!\!\!\!\!\!\!\!\!\!\langle |k_x|^{\alpha}\rangle_{angle}=\frac{\Gamma[(\alpha+1)/2]\Gamma(d/2)}{\Gamma[(d+\alpha)/2]\sqrt{\pi}} |k|^{\alpha}, \end{eqnarray} where we set $\langle |k|^{\alpha}\rangle=|k|^{\alpha}$. We find, 
\begin{eqnarray}&&\!\!\!\!\!\!\!\!\!\!\!\!\!\!
P_0(x)\!\sim \! -\frac{\sqrt{2}K_{\alpha}S_{d-1}}{x^{1+\alpha}(2\pi)^{d/2} \sqrt{\pi}}\int_0^{\infty}\!\! k^{\alpha+(d-1)/2}dk\nonumber\\&&\!\!\!\!\!\!\!\!\!\!\!\!\!\times   \cos\left(k-\frac{\pi (d-1)}{4}\right)\exp\left[-\epsilon k\right],\nonumber
\end{eqnarray}
where we performed the integral over angles. We write, 
\begin{eqnarray}&&\!\!\!\!\!\!\!\!\!\!\!\!\!\!
P_0(x)\sim -\frac{\sqrt{2}K_{\alpha}S_{d-1}}{x^{1+\alpha}(2\pi)^{d/2} \sqrt{\pi}} \text{Re} \left[\exp\left(\frac{i\pi (d-1)}{4}\right)\right.\nonumber\\&&\!\!\!\!\!\!\!\!\!\!\!\!\!\!\left.\times
\int_0^{\infty}\!\!\!\! \exp\left[-(\epsilon+i) k\right]k^{\alpha+(d-1)/2} dk\right].
\end{eqnarray} 
By using integration variable $t=(\epsilon+i)x$ we find
\begin{eqnarray}&&\!\!\!\!\!\!\!\!\!\!\!\!\!\!
P_0(x)\sim -\frac{\sqrt{2}K_{\alpha}S_{d-1}\Gamma(\alpha+(d+1)/2)}{x^{1+\alpha}(2\pi)^{d/2} \sqrt{\pi}}\nonumber\\&&\!\!\!\!\!\!\!\!\!\!\!\!\!\! \text{Re} \left[\exp\left(\frac{i\pi (d-1)}{4}\right)(\epsilon+i)^{-\alpha-(d+1)/2}\right].
\end{eqnarray} 
We conclude that at $x\gg K_{\alpha}^{1/\alpha}$,
\begin{eqnarray}&&\!\!\!\!\!\!\!\!\!\!\!\!\!\!
P_0(x)\sim \frac{\sqrt{2}K_{\alpha}S_{d-1}\Gamma(\alpha+(d+1)/2)\sin(\pi \alpha/2)}{x^{1+\alpha}(2\pi)^{d/2} \sqrt{\pi}}.
\end{eqnarray} 
For the probability density function we find restoring physical units that at $x\gg (K_{\alpha}t)^{1/\alpha}$,
\begin{eqnarray}&& \!\!\!\!\!\!\!\!\!\!\!\!\!\!
P(x, t)\sim \frac{\sqrt{2}K_{\alpha}S_{d-1}\Gamma(\alpha+(d+1)/2)\sin(\pi \alpha/2)t}{x^{1+\alpha}(2\pi)^{d/2} \sqrt{\pi}},\label{loevy}
\end{eqnarray}
where we used Eq.~(\ref{rm}). Other form is obtained using the value of $K_{\alpha}$ given by Eq.~(\ref{difc}),
\begin{eqnarray}&& \!\!\!\!\!\!\!\!\!\!\!\!\!\!\!\!
P(x, t)\!\sim \!\frac{\Gamma[(\alpha+1)/2]\Gamma(\alpha+(d+1)/2)tA\langle |v|^{\alpha}\rangle}{2^{(d-1)/2} |\Gamma(1\!-\!\alpha)|\Gamma(\alpha)\Gamma[(d\!+\!\alpha)/2]\langle \tau\rangle x^{1+\alpha}},\label{o}
\end{eqnarray}
where we used $\Gamma(\alpha)\Gamma(1-\alpha)=\pi/\sin(\pi\alpha)$. In three-dimensional and one-dimensional cases we find,
\begin{eqnarray}&& \!\!\!\!\!\!\!\!\!\!\!\!\!\!
P(x, t)\sim \frac{\alpha tA\langle |v|^{\alpha}\rangle}{|\Gamma(1-\alpha)|\langle \tau\rangle x^{1+\alpha}}, \ \ d=1, 3. \label{o1}
\end{eqnarray}
In one dimension this is known result \cite{godreche,BarkaiPhysLett,BarkaiPhysRev}. The coincidence of tails in one- and three-dimensional cases does not have clear origin. In other dimensions including the physically relevant tow-dimensional case the tail is different and dimension-dependent. 

We conclude that despite that the PDF of $x$ is non-universal, its tail obeys universal power-law with decay exponent $\alpha+1$. Furthermore though the velocity statistics can be non-isotropic the tail is determined uniquely by $\langle v^{\alpha}\rangle$ so effective isotropization occurs. This could look contradicting the non-isotropy of both the bulk and the tail of the PDF demonstrated in previous and coming Sections. In fact isotropization happens only in the leading order term of large $x$ series of $P(x, t)$. It is seen readily from Eq.~(\ref{xp}) that the next order term involves the angular integral $\int d{\hat k} \left\langle |{\hat k}\cdot\bm v|^{\alpha}\right\rangle^2$ that depends on non-isotropy of the velocity statistics (the orders of integration and averaging can no longer be interchanged because of non-linearity). Thus the large $x$ series of $P(x, t)$ is isotropic in leading order only.  

We illustrate the conclusions of this Section with two-dimensional case of $XYZ...$ model, called non-symmetric $x-y$ model. In this model diffusion coefficients in $x$ and $y$ directions differ so diffusion is non-symmetric. The velocity vector is in $x$ or $-x$ direction with probability $\lambda/2$ and in $y$ or $-y$ directions with probability $(1-\lambda)/2$. The magnitude of velocity $v_0$ is fixed. Thus $\langle |\bm k\cdot\bm v|^{\alpha}\rangle=v_0^{\alpha}[\lambda |k_x|^{\alpha}+(1-\lambda)|k_y|^{\alpha}]$. We find from Eq.~(\ref{lw}) that,
\begin{eqnarray}&&\!\!\!\!\!\!\!\!\!\!\!
P(\bm x, t)\sim\frac{1}{(K_x K_y t^2)^{1/\alpha}}L\left(\frac{x}{(K_x t)^{1/\alpha}}\right)L\left(\frac{y}{(K_y t)^{1/\alpha}}\right),\nonumber\\&&\!\!\!\!\!\!\!\!\!\!\!
K_x\!=\!\frac{Av_0^{\alpha}\lambda}{\langle \tau\rangle}\left|\cos\left(\frac{\pi\alpha}{2}\right)\right|,\ \ K_y\!=\!\frac{Av_0^{\alpha}(1\!-\!\lambda)}{\langle \tau\rangle}\left|\cos\left(\frac{\pi\alpha}{2}\right)\right|.\nonumber
\end{eqnarray}
This distribution is non-symmetric where asymmetry results from different scaling factors of one-dimensional distributions in $x$ and $y-$directions. The PDF $P(x, t)$ of the distance $x$ from the origin is (in the previous equation $x$ is $x-$component of $\bm x$, below $x=|\bm x|$ with no ambiguity),  
\begin{eqnarray}&&\!\!\!\!\!\!\!\!\!\!\!
P(x, t) \sim \frac{x}{(K_x K_y t^2)^{1/\alpha}}\int_0^{2\pi} L\left(\frac{x\cos\phi}{(K_x t)^{1/\alpha}}\right)\nonumber\\&&\!\!\!\!\!\!\!\!\!\!\! \times
L\left(\frac{x\sin\phi}{(K_y t)^{1/\alpha}}\right)d\phi.\label{ing}
\end{eqnarray}
This PDF depends on the degree of asymmetry via $\lambda$ included in $K_x$, $K_y$. In the limits of $\lambda\to 0$ or $\lambda\to 1$ it gives LW in the direction of the corresponding axis. When $\lambda=1/2$ we find the distribution of $x-y$ model. In contrast the large $x$ asymptotic form of $P(x, t)$ depends on $\langle |v|^{\alpha}\rangle=v_0^{\alpha}$ only and thus is independent of $\lambda$, see Eq.~(\ref{o}). The simplest way of verifying this seems to be using integral representation of $L(x)$ and repeating the steps of the previous derivation in the general case.


The universal power-law tail obtained above has non-trivial implications for the moments of distance from the origin. The moment of $q-$th order $\langle x^q(t)\rangle$ diverges at large $x$ is $q>\alpha$. Consequently the second moment is determined by the tail of the PDF independently of the statistics of velocity. The moments of order $0<q<\alpha$ are determined by the bulk of the PDF. We presume that the moments are determined either by the bulk or by the tail of the PDF which is usually the case and is confirmed below. Then demanding self-consistency we find the described conclusions. We find from Eqs.~(\ref{rm}), 
\begin{eqnarray}&&\!\!\!\!\!\!\!\!\!\!\!\!\!\!
\langle |x|^q(t)\rangle \sim c_q t^{q/\alpha},\ \ q<\alpha
\end{eqnarray}
where the constants $c_q$ are given by 
\begin{eqnarray}&&\!\!\!\!\!\!\!\!\!\!\!\!\!\!
c_q=\int_0^{\infty}x^qP_0(x)dx,
\end{eqnarray}
where $P_0(x)$ is given by Eq.~(\ref{pi0}). For $q>\alpha$ the integral diverges and the formula fails demanding finding the tail of the PDF. This is obtained in Section \ref{infn}. Before we study the tail of teh distribution we provide consequence of our consideration for sums of random variables. 

\section{Statistics of sum of power-law tailed variables in $d$ dimensions}\label{sm}

In this Section we describe the correspondence between our results and distribution of sum of independent identically distributed random vectors whose PDF has power-law tail with infinite dispersion but finite mean, cf. \cite{UchZol,st}.  We observe that at large times the bulk distribution of $\bm x(t)$ is identical with that of, 
\begin{eqnarray}&&\!\!\!\!\!\!\!\!\!\!\!\!\!\!
\bm x_0(t)=\sum_{i=1}^{N(t)}\bm v_i\tau_i.\label{g2}
\end{eqnarray}
Indeed, comparing this definition with Eq.~(\ref{walkoro}) we see that equating distributions in the bulk is equivalent to neglecting at large times the displacement due to the last step of the walk. This point is obvious for ordinary random walks but not for LWs: it breaks down for ballistic LW with infinite average duration of the step \cite{BarkaiPR}. In our case the proof of the distributions' equality is done using the Montroll-Weiss equation. It is demonstrated in Appendix \ref{mw} that the Montroll-Weiss equation for the variable $\bm x_0(t)$ is very similar to that for $\bm x(t)$. The prefactor that is different obeys the small $u$ behavior, 
\begin{eqnarray}&& \!\!\!\!\!\!\!\!\!\!\!\!\!\! \frac{1-\psi(ut^{-1})}{ut^{-1}}=\langle \tau\rangle+o(t),
\end{eqnarray}
that is identical with that in Eq.~(\ref{lmt}). Thus the long-time limits for the bulk of the PDFs of $\bm x(t)$ and $\bm x_0(t)$ coincide. 

We obtain characteristic function of 
\begin{eqnarray}&& \!\!\!\!\!\!\!\!\!\!\!\!\!\! 
\bm Y_N=\frac{\sum_{i=1}^{N}\bm v_i\tau_i}{N^{1/\alpha}},
\end{eqnarray} 
in $N\to \infty$ limit and then demonstrate that it is equivalent to our previous results. We have, 
\begin{eqnarray}&& \!\!\!\!\!\!\!\!\!\!\!\!\!\! 
P_N(\bm k)=\left\langle \exp[i\bm k \cdot \bm Y_N]\right\rangle=\left\langle \exp\left(\frac{i\bm k \cdot \bm v \tau}{N^{1/\alpha}}\right)\right\rangle^N,
\end{eqnarray}
where we used independence of the summands in $\bm Y_N$. Performing averaging over $\tau$,
\begin{eqnarray}&& \!\!\!\!\!\!\!\!\!\!\!\!\!\! 
P_N(\bm k)=\left\langle \psi\left(u=\frac{\epsilon-i\bm k \cdot \bm v}{N^{1/\alpha}}\right)\right\rangle^N,
\end{eqnarray}
where infinitesimal $\epsilon$ is introduced because the Laplace transform $\psi(u)$ is defined uniquely for complex $u$ with positive real part. We find performing averaging over $\bm v$ in the small argument expansion of $\psi(u)$ given by Eq.~(\ref{smlas}),
\begin{eqnarray}&& \!\!\!\!\!\!\!\!\!\!\!\!\!\! 
\lim_{N\to\infty}P_N(\bm k)=\lim_{N\to\infty}\left(1+\frac{A\langle(\epsilon-i\bm k \cdot \bm v)^{\alpha}\rangle}{N}\right)^N\nonumber\\&&\!\!\!\!\!\!\!\!\!\!\!\!\!\!=\exp\left(-A\left|\cos\left(\frac{\pi \alpha}{2}\right)\right|\langle \left|\bm k \cdot \bm v\right|^{\alpha} \rangle\right),\label{ch}
\end{eqnarray}
where we used Eq.~(\ref{branch}). This result reproduces Eq.~(\ref{lw}) for distribution of $\bm x_0(t)$ using that with probability one, 
\begin{eqnarray}&&\!\!\!\!\!\!\!\!\!\!\!\!\!\!
\frac{\bm x_0(t)\langle \tau\rangle^{1/\alpha}}{t^{1/\alpha}}=\frac{\sum_{i=1}^{N(t)}\bm v_i\tau_i}{N^{1/\alpha}(t)},\label{rndsm}
\end{eqnarray}
where we used the law of large numbers $t/N(t)=\langle \tau\rangle$ and the limit $t\to\infty$ is assumed. Thus the distribution derived in the previous Section is direct consequence of the distribution of sum of many independent random variables. 

\section{Infinite density for anisotropic L\'{e}vy walks}\label{infn}

In this Section we demonstrate that there is the finite limit, 
\begin{eqnarray}&&\!\!\!\!\!\!\!\!\!\!\!\!\!\!
\lim_{t\to\infty}t^{d-1+\alpha}P(t \bm w, t),\ \ w\neq 0. \label{inf}
\end{eqnarray}
In the next Section we demonstrate that this limit provides the description of the tail of the PDF $P(\bm x, t)$ on scales where $|\bm x|\propto t$. 

We introduce ${\tilde P}(\bm x, t)=P(\bm x, t)-\delta(\bm x)$. We have from the Montroll-Weiss equation (\ref{basic}) that the Fourier-Laplace transform ${\tilde P}(\bm k, u)$ of ${\tilde P}(\bm x, t)$ obeys, 
\begin{eqnarray}
&& \!\!\!\!\!\!\!\!\!\!\!\!\!\!{\tilde P}(\bm k, u)\!=\!\left\langle\frac{1\!-\!\psi(u\!-\!i\bm k\cdot\bm v)}{u\!-\!i\bm k\cdot\bm v}\right\rangle\frac{1}{1\!-\!\left\langle \psi(u\!-\!i\bm k\cdot\bm v)\right\rangle}\!-\!\frac{1}{u}.\label{tld}
\end{eqnarray}
We use, 
\begin{eqnarray}&&\!\!\!\!\!\!\!\!\!\!\!\!\!\!
{\tilde P}(t \bm w, t)=\int\frac{d\bm k'}{(2\pi)^d}\frac{du'}{2\pi i}\exp\left[it\bm k'\cdot\bm w+u't\right]{\tilde P}(\bm k', u')\nonumber\\&&\!\!\!\!\!\!\!\!\!\!\!\!\!\!=t^{-d-1}\int\frac{d\bm k}{(2\pi)^d}\frac{du}{2\pi i}\exp\left[i\bm k\cdot\bm w\right]{\tilde P}\left(\frac{\bm k}{t}, \frac{u}{t}\right).
\end{eqnarray}
We prove the existence of finite limit, \begin{eqnarray}&&\!\!\!\!\!\!\!\!\!\!\!\!\!\!
{\tilde P}_i(\bm k, u)=\lim_{t\to\infty}\!t^{\alpha-2}{\tilde P}\left(\frac{\bm k}{t}, \frac{u}{t}\right),
\end{eqnarray}
where $i$ stands for "infinite". We have \begin{eqnarray}&&\!\!\!\!\!\!\!\!\!\!\!\!\!\!
\lim_{t\to\infty}\frac{{\tilde P}(t \bm w, t)}{t^{1-d-\alpha}}\!=\! \int\!\!\!\frac{d\bm k}{(2\pi)^d}\frac{du}{2\pi i}\exp\left[i\bm k\cdot\bm w\!+\!u\right]{\tilde P}_i(\bm k, u).\label{intr}\end{eqnarray}
We use the large $t$ asymptotic forms ($\langle \bm k\cdot\bm v\rangle=0$) for $\psi$ which is Laplace transform of $\psi(\tau)$,
\begin{eqnarray}&& \!\!\!\!\!\!\!\!\!\!\!\!\!\!\left\langle\frac{1\!-\!\psi(ut^{-1}\!-\!i\bm k\cdot\bm vt^{-1})}{ut^{-1}\!-\!i\bm k\cdot\bm vt^{-1}}\right\rangle\!\sim\!\langle \tau\rangle\!-\!A\left\langle \left(
\frac{u}{t}\!-\!\frac{i\bm k\cdot\bm v}{t}\right)^{\alpha-1}\right\rangle,\nonumber
\end{eqnarray}
and 
\begin{eqnarray}&& \!\!\!\!\!\!\!\!\!\!\!\!\!\! \left[1\!-\!\left\langle \psi\left(\frac{u}{t}\!-\!\frac{i\bm k\cdot\bm v}{t}\right)\right\rangle\right]^{-1}\!\sim\!\frac{t}{\langle \tau\rangle u}\!\\&&\!\!\!\!\!\!\!\!\!\!\!\!\!\!+\!\left(\frac{u}{t}\right)^{\alpha-2}\frac{A}{\langle \tau\rangle^2}\left\langle \left(
1\!-\!\frac{i\bm k\cdot\bm v}{u}\right)^{\alpha}\right\rangle,
\end{eqnarray}
holding when the rest of the arguments are held fixed. We neglected higher order terms.

The use of these identities in Eq.~(\ref{tld}) gives, 
\begin{eqnarray}&&\!\!\!\!\!\!\!\!\!\!\!\!\!\!
\frac{u^{2-\alpha}{\tilde P}_i(\bm k, u)}{A/\langle \tau\rangle}\!=\!\left\langle \left(
1\!-\!\frac{i\bm k\cdot\bm v}{u}\right)^{\alpha}\right\rangle\!-\!\left\langle\left(
1\!-\!\frac{i\bm k\cdot\bm v}{u}\right)^{\alpha\!-\!1}\right\rangle.\nonumber
\end{eqnarray}
We find using Eq.~(\ref{intr}) and ${\tilde P}(\bm x, t)=P(\bm x, t)$ for $\bm x\neq 0$, 
\begin{eqnarray}&&\!\!\!\!\!\!\!\!\!\!\!\!\!\!
\lim_{t\to\infty}\frac{P(t \bm w, t)}{t^{1-d-\alpha}}=\frac{A}{\langle\tau\rangle}\int \frac{d\bm k}{(2\pi)^d}\frac{du}{2\pi i}u^{\alpha-2}\exp\left[i\bm k\cdot\bm w+u\right]\nonumber\\&&\!\!\!\!\!\!\!\!\!\!\!\!\!\!\left[\left\langle \left(
1\!-\!\frac{i\bm k\cdot\bm v}{u}\right)^{\alpha}\right\rangle\!-\!\left\langle\left(
1\!-\!\frac{i\bm k\cdot\bm v}{u}\right)^{\alpha-1}\right\rangle\right],\ \ w\neq 0.\label{int}\end{eqnarray} It can be seen readily that in one dimension Eq.~(\ref{int}) reproduces the infinite density obtained in \cite{BarkaiPhysRev}. We can use the way of calculation proposed in that work for finding the integral in Eq.~(\ref{int}). We use series, 
\begin{eqnarray}&&\!\!\!\!\!\!\!\!\!\!\!\!\!\!
(1-x)^{\alpha}-(1-x)^{\alpha-1}=\sum_{n=1}^{\infty}\frac{(-\alpha)_n x^n}{\alpha(n-1)!}\end{eqnarray}
where $(a)_n=\Gamma(a+n)/\Gamma(a)=a(a+1)..(a+n-1)$ is the Pochhammer symbol. We find, 
\begin{eqnarray}&&\!\!\!\!\!\!\!\!\!\!\!\!\!\!
\lim_{t\to\infty}\frac{P(t \bm w, t)}{t^{1\!-\!d\!-\!\alpha}}\!=\!\frac{A}{\langle\tau\rangle}\int \frac{d\bm k}{(2\pi)^d}\frac{du}{2\pi i}u^{\alpha\!-\!2\!-\!2n}\exp\left[i\bm k\!\cdot\!\bm w\!+\!u\right]\nonumber\\&&\!\!\!\!\!\!\!\!\!\!\!\!\!\!
\frac{(-\alpha)_{2n} \left\langle[i\bm k\!\cdot\!\bm v]^{2n}\right\rangle}{\alpha(2n\!-\!1)!}\!= \!\!\frac{A}{ \langle\tau\rangle}\!\!\int\!\!\!\frac{d\bm k}{(2\pi)^d}\!\frac{\exp\left[i\bm k\!\cdot\!\bm w\right]\left\langle G_{\alpha}(\bm k\!\cdot\!\bm v)\right\rangle}{|\Gamma(1-\alpha)|},\label{tr}
\end{eqnarray}
where we defined,
\begin{eqnarray}&&\!\!\!\!\!\!\!\!\!\!\!\!\!\!
G_{\alpha}(y)=\sum_{n=1}^{\infty}\frac{(-1)^n y^{2n}}{(2n-1)!(2n-\alpha)(2n+1-\alpha)},\label{tr1}\end{eqnarray}
using $(-\alpha)_{2n}/\Gamma(2n+2-\alpha)=(2n-\alpha)(2n+1-\alpha)/\Gamma(-\alpha)$. The function $G_{\alpha}(y)$ was introduced in \cite{BarkaiPhysRev} where it was demonstrated that,
\begin{eqnarray}&&\!\!\!\!\!\!\!\!\!\!\!\!\!\!
G(y)=\alpha B_{\alpha}(y)-(\alpha-1) B_{\alpha-1}(y),\label{r1}\\&&\!\!\!\!\!\!\!\!\!\!\!\!\!\! B_{\alpha}(y)=\int_0^1 \frac{\cos(\omega y)-1}{\omega^{1+\alpha}}d\omega.
\end{eqnarray}
Thus for finding the inverse Fourier transform (\ref{tr}) we consider, 
\begin{eqnarray}&&\!\!\!\!\!\!\!\!\!\!\!\!\!\!
{\tilde B}_{\alpha}(\bm w, \bm v)=\int\!\frac{d\bm k}{(2\pi)^d}\exp\left[i\bm k\!\cdot\!\bm w\right]{\tilde B}_{\alpha}(\bm k\cdot\bm v)\nonumber\\&&\!\!\!\!\!\!\!\!\!\!\!\!\!\!=\int\!\frac{d\bm k}{(2\pi)^d}\exp\left[i\bm k\!\cdot\!\bm w\right]\int_0^1 \frac{\cos(\omega \bm k\cdot\bm v)-1}{\omega^{1+\alpha}}d\omega.
\end{eqnarray}
We observe that, 
\begin{eqnarray}&&\!\!\!\!\!\!\!\!\!\!\!\!\!\!
\int\!\!\frac{d\bm k}{(2\pi)^d}\!\exp\left[i\bm k\!\cdot\!\bm w\right]\cos(\omega \bm k\cdot\bm v)\!=\!\frac{\delta(\bm w\!+\!\omega\bm v)\!+\!\delta(\bm w\!-\!\omega\bm v)}{2}.\nonumber
\end{eqnarray}
Thus we find, 
\begin{eqnarray}&&\!\!\!\!\!\!\!\!\!\!\!\!\!\!
{\tilde B}_{\alpha}(\bm w, \bm v)\!=\!\int_0^1 \frac{\delta(\bm w\!+\!\omega v{\hat v})\!+\!\delta(\bm w\!-\!\omega v{\hat v})-2\delta(\bm x)}{2\omega^{1+\alpha}}d\omega.
\end{eqnarray}
where $\bm v=v{\hat v}$. We find using the angular $\delta-$function $\delta_a({\hat n})$ obeying $\delta(\bm w-\bm w')=w^{1-d}\delta(w-w')\delta_a({\hat w}-{\hat w'})$ that for $w\neq 0$,
\begin{eqnarray}&&\!\!\!\!\!\!\!\!\!\!\!\!\!\!
{\tilde B}_{\alpha}(\bm w, \bm v)\!=\!v^{\alpha}\int_0^v \delta(w\!-\!\omega') \frac{\delta_a({\hat w}\!+\!{\hat v})\!+\!\delta_a({\hat w}\!-\!{\hat v})}{2\omega'^{1\!+\!\alpha}w^{d\!-\!1}}d\omega',
\end{eqnarray}
where $\omega'=v\omega$. This gives  that, 
\begin{eqnarray}&&\!\!\!\!\!\!\!\!\!\!\!\!\!\!
{\tilde B}_{\alpha}(\bm w, \bm v)=\frac{\delta_a({\hat w}+{\hat v})+\delta_a({\hat w}-{\hat v})}{2v^{-\alpha}w^{d+\alpha}},\ \ |\bm w|<|\bm v|, \\&&\!\!\!\!\!\!\!\!\!\!\!\!\!\!
{\tilde B}_{\alpha}(\bm w, \bm v)=0,\ \ |\bm w|>|\bm v|. 
\end{eqnarray}
We find using Eqs.~(\ref{tr}),(\ref{r1}) that (we switch $\bm w$ with dimensions of velocity by $\bm v$ in the final formula) for $v\neq 0$,  
\begin{eqnarray}&&\!\!\!\!\!\!\!\!\!\!\!\!\!\!
I(\bm v)=\lim_{t\to\infty}\!\frac{P(t \bm v, t)}{t^{1-d-\alpha}}\!=\frac{A}{v^{d-1}|\Gamma(1-\alpha)|\langle \tau\rangle}\nonumber\\&&\!\!\!\!\!\!\!\!\!\!\!\!\!\!\times\int_{v'>v} F(v'{\hat v})v'^{d-1}d v' \left[\alpha\frac{v'^{\alpha}}{v^{1+\alpha}}-(\alpha-1)\frac{v'^{\alpha-1}}{v^{\alpha}}\right],\label{infdensity}
\end{eqnarray}
where $I(\bm v)$ is the infinite density. We separated the dependence in the velocity's PDF $F(\bm v)$ on the magnitude $v$ and direction ${\hat v}$ and used $F(\bm v)=F(-\bm v)$. This is chief result of our work that provides the infinite density as the statement on the existence of long-time scaling limit of the PDF similar to the central limit theorem or the limit introduced in the previous Section. The point $\bm v=0$ describes the region of not too large $|\bm x|$ where L\'{e}vy distribution describes $P(\bm x, t)$ well.  This region that determines the normalization shrinks in the considered large times' scaling limit to the point $x=0$. 

Infinite density inherits anisotropy of $F(\bm v)$:
all angular harmonics present in the expansion of $F(\bm v)$ in spherical harmonics will be present in the expansion of $I(\bm v)$, see Eq.~(\ref{infdensity}) (disregarding degenerate cases when the integral that provides the corresponding coefficient vanishes). Thus in contrast with the case of isotropic statistics described by Eq.~(\ref{infisotr}) the angular structure is non-isotropic when the velocity statistics is not. Furthermore, though anisotropy of the statistics of the single step of the walk influences the distribution in the bulk, described by the anisotropic L\'{e}vy distribution, for the tail this influence is more immediate.

The infinite density limit (\ref{infdensity}) implies that at large times, 
\begin{eqnarray}&&\!\!\!\!\!\!\!\!\!\!\!\!\!\!
P(\bm x, t)\sim \frac{1}{t^{d+\alpha-1}}I\left(\frac{\bm x}{t}\right),\ \ \bm x\neq 0. \label{infinite}
\end{eqnarray} This is complementary to the other limit implied by Eq.~(\ref{lw10}) , \begin{eqnarray}&&\!\!\!\!\!\!\!\!\!\!\!\!\!\! P_L(\bm x, t)\sim \frac{1}{(K_{\alpha}t)^{d/\alpha}}{\hat L}_d\left(\frac{\bm x}{(K_{\alpha}t)^{1/\alpha}}\right).\label{lv} \end{eqnarray} In fact, both limits characterize different asymptotic regions of the PDF. It is clear from the structure of the scaling limits that L\'{e}vy distribution describes the bulk of the PDF whose scale grows proportionally to $t^{1/\alpha}$ and infinite density describes the tail with $|\bm x|\propto t$. In the next Section we demonstrate that the integer order moments are determined by the infinite density tail of the PDF.  

\section{Integer order moments and dispersion at large times} \label{integer}

In this Section we calculate the long-time limit of the moments of integer order from the Montroll-Weiss equation. The basic result that we derive is that 
\begin{eqnarray}&&\!\!\!\!\!\!\!\!\!\!\!\!\!\!
\langle x_i(t)x_k(t)\rangle=\!\frac{2A  t^{3-\!\alpha}\!}{|\Gamma(1\!-\!\alpha)|(2\!-\!\alpha)(3\!-\!\alpha)\langle \tau\rangle}\langle v_iv_k\rangle.
\end{eqnarray}
The property $F(\bm v)=F(-\bm v)$ implies that $\langle v_iv_k\rangle=0$ for $i\neq k$ so we can write,
\begin{eqnarray}&&\!\!\!\!\!\!\!\!\!\!\!\!\!\!
\langle x_i(t)x_k(t)\rangle=\!\frac{2A  t^{3-\!\alpha}\!}{|\Gamma(1\!-\!\alpha)|(2\!-\!\alpha)(3\!-\!\alpha)\langle \tau\rangle}\langle v_i^2\rangle \delta_{ik}.\label{disp}
\end{eqnarray}

The trace of this equation gives, 
\begin{eqnarray}&&\!\!\!\!\!\!\!\!\!\!\!\!\!\!
\langle x^2(t)\rangle=\!\frac{2A  \langle v^2\rangle}{|\Gamma(1\!-\!\alpha)|(2\!-\!\alpha)(3\!-\!\alpha)\langle \tau\rangle} t^{3-\!\alpha} .\label{iunv}
\end{eqnarray}
This is universal formula that holds in arbitrary dimension for arbitrary (possibly strongly anisotropic) statistics of velocity. In the case of isotropic statistics this reduces to the previously derived Eq.~(\ref{moments}). In one dimension this reproduces the formula of \cite{BarkaiPhysRev}. 

We observe that directly repeating the steps of one-dimensional calculation in \cite{BarkaiPhysRev} with $\bm k\cdot \bm v$ instead of $kv$ we find the asymptotic expansion, 
\begin{eqnarray}&&\!\!\!\!\!\!\!\!\!\!\!\!\!\! P(\bm k, t)\!\sim \!1\!+\!\frac{A}{\langle \tau\rangle}\!\sum_{n=1}^{\infty} \!\frac{\Gamma(2n\!-\!\alpha)(-1)^n  t^{2n\!+\!1\!-\!\alpha}\!\left\langle (\bm k\cdot \bm v)^{2n}\right\rangle\!}{(2n\!-\!1)!|\Gamma(1\!-\!\alpha)|\Gamma(2n\!+\!2\!-\!\alpha)}.\nonumber\end{eqnarray}
which is direct continuation of Eq.~(\ref{istr}) for arbitrary statistics of velocity. For $XYZ...$ model we have $\left\langle (\bm k\cdot \bm v)^{2n}\right\rangle=v_0^{2n}[\sum_{i=1}^d k_i^{2n}]/d$ so that $P(\bm k, t)$ has the structure of $P(\bm k, t)=\sum_{i=1}^d P_1(k_i, t)$ where $P_1(k, t)$ is the corresponding one-dimensional distribution. Thus the inverse Fourier transform is sum of products on one-dimensional distribution $P_1(x_i, t)$ times $\delta-$functions of the rest of coordinates. For instance in three dimensions we have
\begin{eqnarray}&&\!\!\!\!\!\!\!\!\!\!\!\!\!\! 
P(\bm x, t)\sim P_1(x, t)\delta(y)\delta(z)+\delta(x)P_1(y, t)\delta(z)\nonumber\\&&\!\!\!\!\!\!\!\!\!\!\!\!\!\!  +\delta(x)\delta(y)P_1(z, t),\label{decoupl}
\end{eqnarray}
where $P_1(x, t)$ is the one-dimensional distribution studied in \cite{BarkaiPhysRev}. Correspondingly the temporal growth of the moments is as in one dimension. 

Comparing the asymptotic form of $P(\bm k, t)$ with the characteristic function, see Eq.~(\ref{ch}),
\begin{eqnarray}&&\!\!\!\!\!\!\!\!\!\!\!\!\!\! P(\bm k, t)=1+\sum_{n=1}^{\infty} \frac{(-1)^n\left\langle (\bm k\cdot \bm x)^{2n}\right\rangle}{(2n)!},
\end{eqnarray}
we read off the moments. We find,  
\begin{eqnarray}&&\!\!\!\!\!\!\!\!\!\!\!\!\!\!\!\!\!\!
\left\langle \prod_{i=1}^d x_i^{\nu_i}(t)\right\rangle\sim\frac{2nA\left\langle \prod_{i=1}^d v_i^{\nu_i}\right\rangle t^{2n+1-\alpha}}{|\Gamma(1-\alpha)|(2n-\alpha)(2n+1-\alpha)\langle \tau\rangle},\label{mom}
\end{eqnarray}
where $\sum_{i=1}^d\nu_i=2n$. The $n=1$ term gives Eq.~(\ref{disp}). On the level of dispersion there is no qualitative difference between isotropic and anistoropic statistics of velocity. The formula (\ref{iunv}) for $\langle x^2(t)\rangle$ is dimension-independent. This is also true for $2n-$th moment of the magnitude of the distance from the walk's origin $\langle x^{2n}(t)\rangle$,
\begin{eqnarray}&&
\langle x^{2n}(t)\rangle\sim\frac{2nA\langle v^{2n}\rangle t^{2n+1-\alpha}}{|\Gamma(1-\alpha)|(2n-\alpha)(2n+1-\alpha)\langle \tau\rangle},\nonumber
\end{eqnarray}
as can be seen opening brackets in $\langle x^{2n}(t)\rangle=\left\langle \left(\sum_{i=1}^d x_i^2(t)\right)^{n}\right\rangle$ and using Eq.~(\ref{mom}). This reproduces Eq.~(\ref{moments}) in the isotropic case and reproduces the result of \cite{BarkaiPhysRev,BarkaiPhysLett} for $\langle x^{2n}(t)\rangle$ in one dimension. Similarly the growth of one of the components is universal:
\begin{eqnarray}&&\!\!\!\!\!\!\!\!\!\!\!\!\!\!\!\!\!\!
\langle x_i^{2n}(t)\rangle\sim\frac{2nA\langle v_i^{2n}\rangle t^{2n+1-\alpha}}{|\Gamma(1-\alpha)|(2n-\alpha)(2n+1-\alpha)\langle \tau\rangle}.\label{onec}
\end{eqnarray}
It is readily confirmed using Eq.~(\ref{mom}) and the construction of the infinite density in the previous Section that the identity
 \begin{eqnarray}&&\!\!\!\!\!\!\!\!\!\!\!\!\!\!
 \left\langle \prod_{i=1}^d x_i^{\nu_i}(t)\right\rangle \sim \frac{1}{t^{d+\alpha-1}}\int \prod_{i=1}^d x_i^{\nu_i} I\left(\frac{\bm x}{t}\right) d\bm x, \nonumber  
\end{eqnarray}
holds. The calculation can be done by direct transfer of the calculation in one dimension \cite{BarkaiPhysRev} using $\bm k\cdot\bm v$ instead of $kv$ in the derivations. Thus integer order moments are determined by the infinite density. We saw previously that the moments of order higher than $\alpha$ are described by the tail of the PDF. This confirms that the tail of the PDF is indeed described by the infinite density (unless the tail cannot be reconstructed from moments of integer order which is not the case here).  

Previous results on the moments coincided with those in one dimension \cite{BarkaiPhysRev}. The difference from the one-dimensional case holds when cross correlations of different components of $\bm x(t)$ are considered. We consider the difference in the case of fourth-order moments. We have,
\begin{eqnarray}&&\!\!\!\!\!\!\!\!\!
\langle x_{i}^2(t)x_{k}^2(t)\rangle\sim\frac{4A\langle  v_{i}^2v_{k}^2\rangle}{|\Gamma(1-\alpha)|(4-\alpha)(5-\alpha)\langle \tau\rangle}t^{5-\alpha}.\label{fourthorder}
\end{eqnarray}
In $XYZ...$ model $\langle  v_{i}^2v_{k}^2\rangle=0$ for $i\neq k$ so that this formula gives zero. This does not tell that the positive quantity $x_{i}^2(t)x_{k}^2(t)$ has zero average but rather this tells that higher order corrections for the asymptotic calculation at large times is needed. This can be done by direct study of the Montroll-Weiss equation.

\section{Tail of the PDF of anisotropic LW and fractional order moments} \label{frct}

We observe from Eq.~(\ref{infdensity}) that angular structure of the infinite density, and thus of the PDF's tail, reflects directly the angular structure of the velocity PDF. This is the consequence of that the tail is formed by rare long ballistic flights. We find for the angular average,  
\begin{eqnarray}&&\!\!\!\!\!\!\!\!\!\!\!\!\!\!
I_0(v)=\int I(v{\hat v})d{\hat v}=\frac{A}{v^{d-1}|\Gamma(1-\alpha)|\langle \tau\rangle}\int_{v'>v} F(\bm v')d \bm v'\nonumber\\&&\!\!\!\!\!\!\!\!\!\!\!\!\!\!\times \left[\alpha\frac{v'^{\alpha}}{v^{1+\alpha}}-(\alpha-1)\frac{v'^{\alpha-1}}{v^{\alpha}}\right].
\end{eqnarray}
We conclude from Eq.~(\ref{infinite}) that the probability density function of the distance $x$ from the walk's origin obeys, 
\begin{eqnarray}&&\!\!\!\!\!\!\!\!\!\!\!\!\!\!
P(x, t)\!=\!x^{d\!-\!1}\int\!\! P(x{\hat x}, t)d{\hat x}\!\sim\!\frac{x^{d\!-\!1}}{t^{d\!+\!\alpha\!-\!1}}I_0\left(\frac{x}{t}\right)\!=\!\frac{A}{|\Gamma(1\!-\!\alpha)|}\nonumber\\&&\!\!\!\!\!\!\!\!\!\!\!\!\!\! \int_{v'>x/t} \frac{F(\bm v')d \bm v'}{\langle \tau\rangle} \left[\alpha\frac{tv'^{\alpha}}{x^{1+\alpha}}-(\alpha-1)\frac{v'^{\alpha-1}}{x^{\alpha}}\right].
\end{eqnarray}
This has universal asymptotic form at small $x$,
\begin{eqnarray}&&\!\!\!\!\!\!\!\!\!\!\!\!\!\!
P(x, t)\sim\frac{\alpha t A\langle v^{\alpha}\rangle}{|\Gamma(1\!-\!\alpha)|\langle \tau\rangle x^{1+\alpha}},\ \ \frac{tv_c}{x}\ll 1,
\end{eqnarray}
where $v_c$ is characteristic value of velocity. This is independent of dimension and details of statistics of velocity. This coincides with the tail of the L\'{e}vy distribution in one and three-dimensional cases, see Eq.~(\ref{o1}). In one-dimension this was observed in \cite{BarkaiPhysRev}. However in other dimensions there is a multiplicative factor difference between the two asymptotic forms. For $I_0(x)$ we have,
\begin{eqnarray}&&\!\!\!\!\!\!\!\!\!\!\!\!\!\!
I_0(x)\sim\frac{\alpha A\langle v^{\alpha}\rangle}{|\Gamma(1\!-\!\alpha)|\langle \tau\rangle x^{d+\alpha}}.\label{asmp}
\end{eqnarray}
It can be demonstrated that the small $x$ form of the infinite density and the large $x$ form of the L\'{e}vy distribution have to be of the same order. We observe that $P_L(\bm x, t)$ in Eq.~(\ref{lv}) obeys at large $t$ and $\bm v\neq 0$
\begin{eqnarray}&&\!\!\!\!\!\!\!\!\!\!\!\!\!\!
\!\frac{P_L(t \bm v, t)}{t^{1-d-\alpha}}\propto \frac{1}{t^{1-d-\alpha+d/\alpha}}{\hat L}_d\left(t^{1-1/\alpha}\frac{\bm v}{(K_{\alpha})^{1/\alpha}}\right)\propto const,\nonumber
\end{eqnarray}
where we use that at large arguments $L_d(\bm x)\propto x^{-d-\alpha}$. Thus $P_L(\bm x, t)$ would give finite constant contribution in the scaling limit described by the infinite density. This contribution comes from the tail of the L\'{e}vy distribution. Then the consideration performed in one-dimensional case in \cite{BarkaiPhysRev} holds in higher-dimensional case demonstrating that the small $x$ form of the infinite density and the large $x$ form of the L\'{e}vy distribution have to be of the same order.

Based on the infinite density tail we can readily derive the moments of order higher than $\alpha$. We have, 
 \begin{eqnarray}&&\!\!\!\!\!\!\!\!\!\!\!\!\!\!
\langle |x(t)|^{q}\rangle \sim \int \frac{|x|^{q}}{t^{d+\alpha-1}} I\left(\frac{\bm x}{t}\right) d\bm x={\tilde c}_q t^{q+1-\alpha},\label{momento}\\&&\!\!\!\!\!\!\!\!\!\!\!\!\!\!
{\tilde c}_q=\int |x|^{q+d-1} I_0\left(x\right)  dx.  \nonumber  
\end{eqnarray}
This formula holds for $q>\alpha$ where the formula's breakdown at smaller $q$ is signalled by the small $x$ divergence of the integral in ${\tilde c}_q$ see Eq.~(\ref{asmp}).

The scaling exponents of the moments depend on the moment's order linearly both at $q<\alpha$ and $q>\alpha$, albeit with different linear dependencies. This bilinear behavior of the moments of the distance from the origin was observed in the one-dimensional case in \cite{BarkaiPhysRev} and continues to hold in the higher-dimensional case, cf. \cite{Vulpiani}.

We find the infinite density in the uniform model with fixed velocity $v_0$. In this case the PDF of $|\bm v|$ is $\delta(v-v_0)$ so that 
\begin{eqnarray}&&\!\!\!\!\!\!\!\!\!\!\!\!\!\!
I(x)\!=\!\frac{A v_0^{\alpha-1}(\alpha\!-\!1)}{|\Gamma(1\!-\!\alpha)|\tau x^{\alpha}}\left[\frac{\alpha v_0}{x (\alpha\!-\!1)}\!-\!1\right],\ \ x<v_0,
\end{eqnarray}
If $x>v_0$ then $I(x)=0$. Since $x_c(t)/t\propto t^{(1-\alpha)/\alpha}$ is decaying function of time then at large times the infinite density describes non-trivial region of the the probability density function. We find that if $x_c(t)<x<v_0 t$ then, 
\begin{eqnarray}&&\!\!\!\!\!\!\!\!\!\!\!\!\!\!
P_d(\bm x, t)\sim \frac{A\Gamma(d/2)v_0^{\alpha-1}(\alpha\!-\!1)}{2\pi^{d/2}x^{d+\alpha-1}|\Gamma(1\!-\!\alpha)|\tau}\left[\frac{\alpha v_0 t}{x(\alpha\!-\!1)}\!-\!1 \right]. \label{infi}
\end{eqnarray}
The probability density function at $x>v_0 t$ vanishes since the particle moving at constant speed $v_0$ cannot pass distances longer than $v_0 t$ in time $t$.

\section{Conclusions}

We obtained the PDF $P(\bm x, t)$ for $d-$ dimensional L\'{e}vy walks. We derived two complimentary limiting distributions describing the PDF at long times. 

One of these scalings has origin similar to the central limit theorem providing the $d-$dimensional counterpart of 
one-dimensional L\'{e}vy distribution that describes the bulk of the PDF. There is however significant difference from the 
one-dimensional case. In one dimension different statistics of velocity result in the same universal (up to rescaling of the density) shape 
of the bulk of the PDF $P(x, t)$. This is no longer true in higher dimensions: 
different velocity  statistics result in different shapes of the PDF bulks. Only in the case of  isotropic statistics a certain  universality holds. 
This is in sharp contrast with the normal diffusion, where the $d-$dimensional  Gaussian profiles are universal attractors.

We demonstrated that despite the non-universality of the PDF in the bulk region, the tail of PDF of the distance follows the universal power-law with exponent $-1-\alpha$. The prefactor of the law depends only on angle-averaged statistics of velocity via $\langle |v|^{\alpha}\rangle$ (in contrast the bulk of the PDF depends on probabilities of moving in different directions). That provides direct generalization of the one-dimensional result. Thus dependence on anisotropy of velocity statistics and non-universality of the shape of the PDF in the bulk disappear at large distances where universality is restored. The higher order corrections are anisotropic and non-universal though. 

This universality of the tail's exponent guarantees that, independently of details of anisotropy of the statistics, the moments of order smaller than $\alpha$ where $1<\alpha<2$ are determined by the PDF's bulk. However higher order moments, including dispersion, are due to the PDF's tail. 

The existence of the complimentary scaling, unfamiliar in the field until very recently, is the consequence of the scaling of the tail of the 
PDF $\psi(\tau)$ of the single step duration $\tau$. This limit describes the tail of $P(\bm x, t)$ formed by ballistic motions with very large duration $\tau$. It describes $x$ that scale proportional with $t$ (ballistic scaling). In contrast the displacements described by the bulk of the PDF are formed by accumulation of lots of typical diffusive increments. Thus the bulk describes $x$ that have (anomalous) diffusive scaling proportional to $t^{1/\alpha}$ (we remark that $t^{1/\alpha}\ll t$ at large $t$ because of $\alpha>1$). 

The function that describes the PDF's tail is called the 'infinite density' 
because it is not normalizable. That is caused by divergence of normalization at small arguments. 
It is important to understand that 'infinite density' is not a probability density function (which must be normalized).
The infinite density is a point-like limit of the rescaled PDF and does not have to be normalizable. This is because for different spatial arguments the pointwise 
limit holds at different times. No matter how large time is, the infinite density never converges to the 
PDF uniformly in space. For any large but finite time the bulk of the PDF that determines 
the normalization holds around the origin $\bm x = 0$ and can be described by the corresponding anisotropic L\'{e}vy distribution. 
The complete picture of the PDF is simpler in three and one-dimensional cases. There the small argument form of 
the infinite density tail continuously transforms in the universal tail of the PDF's bulk. This makes it plausible that intermediate region between the bulk and the tail is described by these asymptotic forms. Both the bulk anisotropic L\'{e}vy distribution and its tail shrink to zero when rescaled with ballistic scaling $t$ of the infinite density scaling 
limit. In the two-dimensional case the intermediate region between the bulk and the tail demands separate study.  


\textbf{ Acknowledgement}. This work was supported by the
Russian Science Foundation grant No.\ 16-12-10496 (SD and VZ).
IF and EB thank the Israel Science Foundation for the support. 

{}
\begin{appendices} \appendix

\section{Montroll-Weiss equation from sum over trajectories}\label{mw}

In this Section we derive Montroll-Weiss (MW) equation by using the Laplace transform 
of the characteristic function of the coordinate of random walker. 
Though this equation is  not solvable in general case, 
it does help in evaluating the long-time asymptotic form of the PDF $P(\bm x,t)$. 
The MW-equation is well-known in one-dimensional setup, 
see e. g. \cite{Zaburdaev} and references therein. 
Here we provide alternative derivation in terms of sums over all possible trajectories. 
This can help in the study of multi-time statistics of the walk, fluctuations, and other quantities.

Trajectories during  the period $[0, t]$ can be characterized by the integer number $N\geq 0$ 
of time renewals that occurred during this time interval. 
Since events with different $N$ form a complete set of non-overlapping  events, 
we can write averages as sums of contributions of events with different $N$. This is realized by using identity
\begin{eqnarray}&& \!\!\!\!\!\!
1= \sum_{N=1}^{\infty}  \int_0^t dt' \!\int_{t-t'}^{\infty}dt'' \! \delta\left(\sum_{i=1}^N \tau_i-t'\right)\delta\left(\tau_{N+1}-t''\right)\nonumber\\&&+\int_t^{\infty} \delta(\tau_1-t')dt',\label{one}
\end{eqnarray}
that holds for arbitrary infinite sequence of positive numbers $\tau_i$ and $t$. Only one term on the RHS is non-zero and is one. This is the term of $N$ renewals for which $\leq \sum_{i=1}^N \tau_i< t<  \sum_{i=1}^{N+1} \tau_i$ or the last term of no renewals if $\tau_1>t$. The identity can be used for averaging arbitrary function of $\tau_i$ as sum of contributions of events with different $N$. The simplest is the probability $P_N(t)$ of having $N$ renewals before the time $t$ is, 
\begin{eqnarray}&&\!\!\!\!\!\!\!\!\!\!\!\!
P_N(t)=\left\langle \int_0^t dt' \!\int_{t-t'}^{\infty}dt'' \! \delta\left(\sum_{i=1}^N \tau_i-t'\right)\delta\left(\tau_{N+1}-t''\right)\right\rangle\nonumber\\&&\!\!\!\!\!\!\!\!\!\!\!\!=
\int_0^t p_N(t')dt' \int_{t-t'}^{\infty} \psi(t'')dt'', \nonumber\\&&\!\!\!\!\!\!\!\!\!\!\!\! P_0(t)=\left\langle \int_t^{\infty} \delta(\tau_1-t')dt'\right\rangle=\int_{t}^{\infty} \psi(t')dt',\label{basic1}
\end{eqnarray}
where $p_N(t)=\left\langle \delta\left(\sum_{i=1}^N \tau_i-t\right)\right\rangle$ is the PDF of $\sum_{i=1}^N \tau_i$.  Laplace transform of $p_N(t)$ obeys,
\begin{eqnarray}&&
p_N(u)=\int_0^{\infty}\exp[-u t] p_N(t)dt=\left\langle \exp\left(-u\sum_{i=1}^N \tau_i\right)\right\rangle\nonumber\\&&=\psi^N(u),
\end{eqnarray}
where $\psi(u)$ is the Laplace transform of $\psi(\tau)$.  The Laplace transform of convolution in Eq.~(\ref{basic1}) gives,
\begin{eqnarray}&&
P_N(u)=\frac{\psi^N(u)[1-\psi(u)]}{u},\label{lap}
\end{eqnarray}
where we used $\int_0^{\infty}\psi(\tau)d\tau=1$. This formula includes $N=0$ case implying that the Laplace transform of the generating function $p(s, t)=\sum s^NP_N(t)$ is ($|s|\leq 1$)
\begin{eqnarray}&&\!\!\!\!\!\!\!\!\!
p(s, u)=\frac{[1-\psi(u)]}{u\left[1-s\psi(u)\right]},
\end{eqnarray}
where we use that $\psi(u)\leq 1$ with equality only at $u=0$. The normalization condition $\sum_{N=0}^{\infty} P_N(t)=1$ implying $\sum_{N=0}^{\infty} P_N(u)=1/u$ is obeyed. The Laplace transform of  the average number of renewals $\langle N(t)\rangle=\sum NP_N(t)$ is,
\begin{eqnarray}&&\!\!\!\!\!\!\!\!\!\!\!\!\!\!\!\!\!\!
\langle N(u)\rangle\!=\!\sum_{N=0}^{\infty} N P_N(u)\!=\!\sum_{N=1}^{\infty} \frac{\psi^N(u)}{u}\!=\!\frac{\psi(u)}{u[1\!-\!\psi(u)]},
\end{eqnarray}
which can be obtained also from $\langle N(u)\rangle=\nabla_sp(s=1, u)$. We compare the behavior of $\langle N(t)\rangle$ in cases with finite and infinite dispersion of $\tau$. If dispersion is finite then the small $u$ behavior of $\psi(u)$ is described by $\psi(u)\sim 1-\langle\tau\rangle u+\langle \tau^2\rangle u^2/2$. This gives, 
\begin{eqnarray}&&\!\!\!\!\!\!\!\!\!\!\!\!\!\!\!\!\!\!
\langle N(u)\rangle\sim \frac{1}{\langle\tau\rangle  u^2}+\frac{\sigma^2-\langle \tau\rangle^2}{2u\langle\tau\rangle^2}+o(u).
\end{eqnarray}
where $\sigma^2=\langle \tau^2\rangle-\langle \tau\rangle^2$. We find the long-time behavior,
\begin{eqnarray}&&\!\!\!\!\!\!\!\!\!\!\!\!\!\!\!\!\!\!
\langle N(t)\rangle\sim \frac{t}{\langle\tau\rangle}+\frac{\sigma^2-\langle \tau\rangle^2}{2\langle\tau\rangle^2},
\end{eqnarray}
see e. g. \cite{nebr}. Thus the leading order correction to the law of large number is constant. In the case of divergent $\langle \tau^2\rangle$ the correction grows with time. Using the small $u$ behavior $\psi(u)\sim 1-\langle\tau\rangle u+A u^{\alpha}$ we have 
\begin{eqnarray}&&\!\!\!\!\!\!\!\!\!\!\!\!\!\!\!\!\!\!
\langle N(u)\rangle\!=\frac{1-\langle\tau\rangle u+A u^{\alpha}}{u^2[\langle\tau\rangle -A u^{\alpha-1}]}\sim \frac{1}{\langle\tau\rangle  u^2}+\frac{Au^{\alpha-3}}{\langle\tau\rangle^2}.
\end{eqnarray}
This yields the long-time asymptotic behavior,
\begin{eqnarray}&&\!\!\!\!\!\!\!\!\!\!\!\!\!\!\!\!\!\!
\langle N(t)\rangle\sim \frac{t}{\langle\tau\rangle}+\frac{At^{2-\alpha}}{\langle\tau\rangle^2\Gamma(3-\alpha)}.
\end{eqnarray}
This indicates that convergence of $N(t)/t$ to its long-time probability one limit $1/\langle\tau\rangle$ is slower than in the case of finite $\langle\tau^2\rangle$ because the average of $N(t)/t-1/\langle\tau\rangle$ decays as $t^{1-\alpha}$ which is slower than $1/t$ law of finite dispersion. 

Similarly for $\langle N^2\rangle (u)$ we find, 
\begin{eqnarray}&&\!\!\!\!\!\!\!\!\!\!\!\!\!\!\!\!\!\!
\langle N^2 (u)\rangle \!-\!\langle N(u)\rangle\!=\!\nabla_s^2p(s=1, u)=\frac{2\psi^2(u)}{u\left[1\!-\!\psi(u)\right]^2}.
\end{eqnarray}
In the case of finite $\langle \tau^2\rangle$ this has small $u$ behavior, 
\begin{eqnarray}&&\!\!\!\!\!\!\!\!\!\!\!\!\!\!\!\!\!\!
\langle N^2 (u)\rangle \!-\!\langle N(u)\rangle\sim\frac{2(1-2\langle\tau\rangle u)}{\langle\tau\rangle^2 u^3[1-\langle \tau^2\rangle u/\langle\tau\rangle]}\nonumber\\&&\!\!\!\!\!\!\!\!\!\!\!\!\!\!\!\!\!\! 
\sim \frac{2}{\langle\tau\rangle^2 u^3}-\frac{4}{u^2\langle\tau\rangle}+\frac{2\langle\tau^2\rangle}{u^2\langle\tau\rangle^3}.
\end{eqnarray}
This gives the long-time behavior, 
\begin{eqnarray}&&\!\!\!\!\!\!\!\!\!\!\!\!\!\!\!\!\!\!
\langle N^2 (t)\rangle -\langle N(t)\rangle\sim\frac{t^2}{\langle\tau\rangle^2}-\frac{4t}{\langle\tau\rangle}+\frac{2t\langle\tau^2\rangle}{\langle\tau\rangle^3}.
\end{eqnarray}
Using the previous result for $\langle N(t)\rangle$ we find for variance, 
\begin{eqnarray}&&\!\!\!\!\!\!\!\!\!\!\!\!\!\!\!\!\!\!
\langle N^2 (t)\rangle -\langle N(t)\rangle^2\sim \frac{\sigma^2t}{\langle\tau\rangle^3}.
\end{eqnarray}
Thus in the case of finite dispersion of $\tau$ variance of $N(t)$ equals the mean up to multiplicative constant. 

In contras in the case of infinite $\langle \tau^2\rangle$ strong violation of Possonicity holds. Using $\psi(u)\sim 1-\langle\tau\rangle u+A u^{\alpha}$ we have,
\begin{eqnarray}&&\!\!\!\!\!\!\!\!\!\!\!\!\!\!\!\!\!\!
\langle N^2 (u)\rangle \!-\!\langle N(u)\rangle\sim\frac{2(1-2\langle\tau\rangle u)}{\langle\tau\rangle^2 u^3[1-2A u^{\alpha-1}/\langle\tau\rangle]}\nonumber\\&&\!\!\!\!\!\!\!\!\!\!\!\!\!\!\!\!\!\! 
\sim \frac{2}{\langle\tau\rangle^2 u^3}+\frac{4Au^{\alpha-4}}{\langle\tau\rangle^3},
\end{eqnarray}
which gives the long-time behavior, 
\begin{eqnarray}&&\!\!\!\!\!\!\!\!\!\!\!\!\!\!\!\!\!\!
\langle N^2 (t)\rangle -\langle N(t)\rangle\sim\frac{t^2}{\langle\tau\rangle^2}+\frac{4At^{3-\alpha}}{\langle\tau\rangle^3\Gamma(4-\alpha)}.
\end{eqnarray}
We find for variance, 
\begin{eqnarray}&&\!\!\!\!\!\!\!\!\!\!\!\!\!\!\!\!\!\!
\langle N^2 (t)\rangle -\langle N(t)\rangle^2\sim \frac{2(\alpha-1)At^{3-\alpha}}{\langle\tau\rangle^3\Gamma(4-\alpha)},
\end{eqnarray}
where $\langle N(t)\rangle$ is small correction to the RHS. We have super-Poissonian behavior, 
\begin{eqnarray}&&\!\!\!\!\!\!\!\!\!\!\!\!\!\!\!\!\!\!
\frac{\langle N^2 (t)\rangle -\langle N(t)\rangle^2}{\langle N(t)\rangle}\sim \frac{2(\alpha-1)At^{2-\alpha}}{\langle\tau\rangle^3\Gamma(4-\alpha)}.
\end{eqnarray}
The PDF that is of interest for us here is that of the particle's displacement $\bm x(t)$ in time $t$. If $N$ renewals occurred in time $t$ then the displacement $\bm x_N(t)$ obeys, 
\begin{eqnarray}&&\!\!\!\!\!\!\!\!\!\!\!\!\!\!\!\!\!\!
\bm x_N(t)=\sum_{i=1}^N \bm v_i \tau_i+\bm v_{N+1}\left(t-\sum_{i=1}^N \tau_i\right),\ \ \bm x_0=\bm v_1 t.\label{our}
\end{eqnarray}
Inserting $1$ in $P(\bm x, t)=\langle 1\times\delta(\bm x(t)-\bm x)\rangle$ in the form given by Eq.~(\ref{one}) we find for the PDF of the particle's position that 
\begin{eqnarray}
&& P(\bm x, t)=\langle \delta(\bm x(t)-\bm x)\rangle=\int_t^{\infty}\langle \delta(\bm v_1 t-\bm x) \delta(\tau_1-t')\rangle dt'\nonumber\\&&+ 
\sum_{N=1}^{\infty} \left\langle \int_0^t dt' \!\int_{t-t'}^{\infty}dt'' \! \delta\left(\sum_{i=1}^N \tau_i-t'\right)\delta\left(\tau_{N+1}-t''\right) \right. \nonumber\\&&\left.
\delta\left(\sum_{i=1}^N \bm v_i \tau_i+\bm v_{N+1}(t-t')-\bm x\right)\right\rangle. \label{dec}
\end{eqnarray}
The characteristic function $P(\bm k, t)=\langle \exp[i\bm k\cdot\bm x(t)]\rangle$ obeys
\begin{eqnarray}
&& P(\bm k, t)=\int_t^{\infty}\langle \exp[i\bm k\cdot \bm v_1 t]  \delta(\tau_1-t')\rangle dt'\nonumber\\&&+ 
\sum_{N=1}^{\infty} \left\langle \int_0^t dt' \!\int_{t-t'}^{\infty}dt'' \! \delta\left(\sum_{i=1}^N \tau_i-t'\right)\psi\left(t''\right) \right. \nonumber\\&&\left.
\exp\left[\sum_{i=1}^N i\bm k\cdot\bm v_i\tau_i+i\bm k\cdot\bm v_{N+1}(t-t')
\right]\right\rangle.
\end{eqnarray}
The Laplace transform over $t$ gives the Montroll-Weiss equation in $d$ dimensions,
\begin{eqnarray}
&& P(\bm k, u)=\left\langle\frac{1-\psi(u-i\bm k\cdot\bm v)}{u-i\bm k\cdot\bm v}\right\rangle\nonumber\\&&+ 
\sum_{N=1}^{\infty} \left\langle\frac{1-\psi(u-i\bm k\cdot\bm v)}{u-i\bm k\cdot\bm v}\right\rangle \left\langle \psi(u-i\bm k\cdot\bm v)\right\rangle^N\nonumber\\&&
=\left\langle\frac{1-\psi(u-i\bm k\cdot\bm v)}{u-i\bm k\cdot\bm v}\right\rangle\frac{1}{1-\left\langle \psi(u-i\bm k\cdot\bm v)\right\rangle}.\label{basica}
\end{eqnarray}
The derivations above hold irrespective of the form of the PDFs of flight times and velocities.

The obtained Montroll-Weiss equation differs from that for the so-called jump model where the particle does not move between renewal times so its coordinate obeys, 
\begin{eqnarray}&&\!\!\!\!\!\!\!\!\!\!\!\!\!\!\!\!\!\!
\bm x_N(t)=\sum_{i=1}^N \bm v_i \tau_i,\ \ \bm x_0=0.\label{jmp}
\end{eqnarray}
This model is more similar to the traditional formulation of displacement as sum of large number of independent random variables that holds for ordinary random walks with constant $\tau$. Repeating the steps of the derivation we find that the Montroll-Weiss equation in this case is, 
\begin{eqnarray}
&& P(\bm k, u)=\frac{1-\psi(u)}{u}\frac{1}{1-\left\langle \psi(u-i\bm k\cdot\bm v)\right\rangle}.
\end{eqnarray}
Statistics of random walks given by Eqs.~(\ref{our}) and (\ref{jmp}) are different though as we demonstrate in the main text are identical in the bulk. 

\section{Small-argument expansion for logarithm of radially symmetric L\'{e}vy distribution}

Here we consider the small argument behavior of $L_d(x)$ defined by Eq.~(\ref{symbulk}). Since $L_d(x)$ is positive function with maximum at $x=0$ it can be advantageous having Taylor expansion for $\ln L_d(x)$ rather than $L_d(x)$ itself [which is provided by Eq.~(\ref{lvtaylor})]. We perform asymptotic study of $L_d(\bm r)$ in Eq.~(\ref{symbulk}). The neighbourhood of the maximum of $L_d(\bm r)$ that holds at $r=0$ can be described writing, 
\begin{eqnarray}&&\!\!\!\!\!\!\!\!\!\!\!\!\!\!
\ln L_d(\bm r)=\ln \left[\langle\exp[i\bm k\cdot \bm r]\rangle_{\bm k} \int \exp[-k^{\alpha}]\frac{d\bm k}{(2\pi)^d}\right].
\end{eqnarray}
where we defined  
\begin{eqnarray}&&\!\!\!\!\!\!\!\!\!\!\!\!\!\! 
\langle\exp[i\bm k\cdot \bm r]\rangle_{\bm k}=\frac{\int \exp[i\bm k\cdot\bm r-k^{\alpha}]d\bm k}{\int \exp[-k^{\alpha}]d\bm k},
\end{eqnarray}
that can be considered as average over the statistics of $\bm k$ defined by the probability density function $P(\bm k)$,
\begin{eqnarray}&&\!\!\!\!\!\!\!\!\!\!\!\!\!\! \langle\exp[i\bm k\cdot \bm r]\rangle_{\bm k}=\!\int \exp[i\bm k\cdot\bm r]P(\bm k)d\bm k,\\&&\!\!\!\!\!\!\!\!\!\!\!\!\!\! P(\bm k)=
\frac{\exp[-k^{\alpha}]}{\int \exp[-k^{\alpha}]d\bm k}=\frac{\alpha\Gamma(d/2)}{2\pi^{d/2}\Gamma(d/\alpha)}\exp[-k^{\alpha}].
\end{eqnarray}
The writing of the integral as average over a statistical distribution is useful because we can use the cumulant expansion theorem for writing $\ln \langle\exp[i\bm k\cdot \bm r]\rangle_{\bm k}$ as series in $\bm r$. We find  
\begin{eqnarray}&&\!\!\!\!\!\!\!\!\!\!\!\!\!\!
L_d(\bm r)= \frac{2^{1-d} \Gamma(d/\alpha)}{\pi^{d/2}\alpha\Gamma(d/2)}\exp\left[\sum_{n=1}^{\infty}\frac{(-1)^n\langle\left(\bm k\cdot \bm r\right)^{2n}\rangle_{c, \bm k}}{(2n)!}\right],
\end{eqnarray}
where $c$ stands for cumulant (see below) and odd-order moments vanish because $P(\bm k)=P(-\bm k)$. Quadratic cumulant is the dispersion, 
\begin{eqnarray}&&\!\!\!\!\!\!\!\!\!\!\!\!\!\!
\langle\left(\bm k\cdot \bm r\right)^{2}\rangle_{c, \bm k}\!=\!r_ir_k \langle k_i k_k\rangle\!=\!\frac{r^2}{d}\langle k^2\rangle\!=\!\frac{r^2\Gamma[(d+2)/\alpha]}{d \Gamma(d/\alpha)}.
\end{eqnarray}
The next non-vanishing term in the series is the quartic cumulant, 
\begin{eqnarray}&&\!\!\!\!\!\!\!\!\!\!\!\!\!\!
\langle\left(\bm k\cdot \bm r\right)^{4}\rangle_{c, \bm k}\!=\!\langle\left(\bm k\cdot \bm r\right)^{4}\rangle-3\langle\left(\bm k\cdot \bm r\right)^{2}\rangle^2\nonumber\\&&\!\!\!\!\!\!\!\!\!\!\!\!\!\!=\frac{3r^4}{d\Gamma(d/\alpha)}\left(\frac{\Gamma[(d+4)/\alpha]}{d+2}-\frac{\Gamma^2[(d+2)/\alpha]}{d \Gamma(d/\alpha)}\right),
\end{eqnarray}
where we used isotropy,
\begin{eqnarray}&&\!\!\!\!\!\!\!\!\!\!\!\!\!\!
\langle\left(\bm k\cdot \bm r\right)^{4}\rangle=r_ir_kr_pr_s \left[\delta_{ik}\delta_{ps}+\delta_{ip}\delta_{ks}+\delta_{is}\delta_{pk}\right]\frac{\langle k^4\rangle}{d(d+2)}\nonumber\\&&\!\!\!\!\!\!\!\!\!\!\!\!\!\!=\frac{3r^4\langle k^4\rangle}{d(d+2)}.
\end{eqnarray}
The series is useful for studying the vicinity of the maximum when $r\ll 1$. Higher $r$ demand higher order cumulants whose form is quite cumbersome. 

The cumulant expansion described above corresponds to resummation of Taylor series for $L_d(\bm r)$ for $\ln L_d(\bm r)$. Both series are useful when $r\ll 1$ is considered. 

\section{Dispersion and fourth order moments at all times}\label{times}

Here we find formulas for dispersion and fourth order moments of the particle's position valid at all times. These are obtained using differentiation of the Montroll-Weiss equation over $\bm k$ and setting $\bm k=0$. This direct procedure brings formulas that hold at arbitrary times in contrast with the asymptotic study in the main text that holds at large times. Thus we find more detailed information on the temporal growth of the moments. This includes corrections to the long-time behavior described in the main text that can in some cases become dominant because of the vanishing of the leading order term. This direct calculation is getting cumbersome for higher order moments so we perform calculation of dispersion and fourth-order moments only. Thus we describe the difference in temporal behavior of $\langle x_{i}^2(t)x_{k}^2(t)\rangle$ for $i\neq k$ in isotropic and $XYZ...$ models.

The displacement's dispersion is found from, 
\begin{eqnarray}&& \!\!\!\!\!\!\!\!\!\!\! \langle x_i^2\rangle\!=\!-\frac{\partial^2}{\partial k_i^2}\left[\left\langle\frac{1\!-\!\psi(u\!-\!i\bm k\cdot\bm v)}{u\!-\!i\bm k\cdot\bm v}\right\rangle\frac{1}{1\!-\!\left\langle \psi(u\!-\!i\bm k\cdot\bm v)\right\rangle}\right],\nonumber \end{eqnarray} 
where the RHS is taken at $k=0$. We find using that odd moments of $v_i$ vanish that 
\begin{eqnarray}&&
\langle x_i^2\rangle=\frac{\left\langle v_i^2\right\rangle}{1-\psi(u)}\left(\frac{1-\psi(u)}{u}\right)''+\frac{\left\langle  v_i^2\right\rangle \psi''(u)}{u [1-\psi(u)]}\nonumber\\&&
=\frac{2\left\langle v_i^2\right\rangle}{u^3}+\frac{2\left\langle v_i^2\right\rangle \psi'(u)}{u^2[1-\psi(u)]}.
\end{eqnarray} 
We find summing over $i$ that, 
\begin{eqnarray}&&\!\!\!\!\!\!\!\!\!\!\!\!\!
\langle x^2(t)\rangle\!=\!2\left\langle v^2\right\rangle \int_{\epsilon-i\infty}^{\epsilon+i\infty}\!\!\frac{\exp[ut]du}{2\pi i} \frac{u \psi'(u)-\psi(u)+1}{u^3[1-\psi(u)]},
\label{dispersion10}\end{eqnarray} 
which can be used for finding detailed temporal dependence of the displacement for given $\psi(\tau)$. This formula is identical for all statistics of velocity. The leading order term in the limit of large times can be obtained using the small $u$-expansion
$\psi(u)\sim 1-\langle \tau\rangle u + Au^{\alpha}$. We have
\begin{eqnarray}&&
u \psi'(u)-\psi(u)+1\sim A(\alpha-1) u^{\alpha},\\&& \frac{u \psi'(u)-\psi(u)+1}{u^3[1-\psi(u)]}={\tilde A}u^{\alpha-4}.\nonumber\end{eqnarray} 
Further,
\begin{eqnarray}&&
\frac{\psi'(u)}{\psi(u)-1}\sim \frac{-\langle \tau\rangle +A\alpha u^{\alpha-1}}{-\langle \tau\rangle u + Au^{\alpha}}=\frac{1}{u[1-Au^{\alpha-1}/\langle \tau\rangle]}
\nonumber\\&&
-\frac{A\alpha u^{\alpha-2}}{\langle \tau\rangle}\approx \frac{1}{u}-\frac{A(\alpha-1) u^{\alpha-2}}{\langle \tau\rangle}.\nonumber
\end{eqnarray} 
We conclude that the leading order term at small $u$ is 
\begin{eqnarray}&&
\frac{\langle x_i^2(u)\rangle}{\left\langle v_i^2\right\rangle}=\frac{\langle r^2(u)\rangle}{\left\langle v^2\right\rangle}\sim\frac{2 A(\alpha-1) u^{\alpha-4}}{\langle \tau\rangle}.\nonumber
\end{eqnarray} 
Thus we find in the $t\to\infty$ limit,
\begin{eqnarray}&&
\langle x^2(t)\rangle\sim \frac{2A\left\langle v^2\right\rangle t^{3-\alpha}}{|\Gamma(1-\alpha)|(2-\alpha)(3-\alpha)\langle \tau\rangle},\nonumber
\end{eqnarray} 
This reproduces the result of the main text directly from Eq.~(\ref{dispersion10}). 
Similar conclusion is reached for 
\begin{eqnarray}&& \!\!\!\!\!\!\!\!\!
\langle x_i^4\rangle=\frac{\partial^4}{\partial k_i^4}\left[\left\langle\frac{1-\psi(u-i\bm k\cdot\bm v)}{u-i\bm k\cdot\bm v}\right\rangle\frac{1}{1-\left\langle \psi(u-i\bm k\cdot\bm v)\right\rangle}\right],\nonumber
\end{eqnarray} 
where the RHS is taken at $k=0$. We find using that odd moments of $v_i$ vanish that, 
\begin{eqnarray}&& \!\!\!\!\!\!\!\!\!
\langle x_i^4\rangle=\frac{\left\langle v_i^4\right\rangle}{1-\psi(u)}\left(\frac{1-\psi(u)}{u}\right)^{(4)}+\frac{6\langle v_i^2\rangle^2 \psi''(u)}{[1-\psi(u)]^2} \nonumber\\&&\!\!\!\!\!\!\!\!\!\times \left(\frac{1-\psi(u)}{u}\right)''+\frac{\left\langle v_i^4\right\rangle \psi^{(4)}(u)}{u[1-\psi(u)]}+\frac{6\langle v_i^2\rangle^2[\psi''(u)]^2}{u[1-\psi(u)]^2}.
\end{eqnarray} 
We observe that at small $u$ we have 
\begin{eqnarray}&& \!\!\!\!\!\!\!\!\!\!\!\!
\frac{1-\psi(u)}{u}=\langle \tau\rangle-Au^{\alpha-1},\ \ 1-\psi(u)=\langle \tau\rangle u-Au^{\alpha},
\end{eqnarray} 
so that 
\begin{eqnarray}&& \!\!\!\!\!\!\!\!\!
\langle x_i^4(u)\rangle\sim \frac{4 A\left\langle v_i^4\right\rangle (-\alpha)_4u^{\alpha-6}}{\langle \tau\rangle\alpha},
\end{eqnarray} 
where the neglected terms involving $\langle v_i^2\rangle^2$ are proportional to $u^{2\alpha-7}$ and can be neglected at small $u$ because of $\alpha>1$. This reproduces the result of the main text,
\begin{eqnarray}&& \!\!\!\!\!\!\!\!\!
\langle x_i^4(t)\rangle\sim \frac{4A\left\langle v_i^4\right\rangle}{|\Gamma(1-\alpha)|(4-\alpha)(5-\alpha)\langle \tau\rangle}t^{5-\alpha}.
\end{eqnarray} 
We consider the cross-correlation $\langle x_i^2x_k^2\rangle$. We have 
\begin{eqnarray}&& \!\!\!\!\!\!\!\!\!
\langle x_i^2(t)x_k^2(t)\rangle=\frac{\partial^4}{\partial k_i^2 \partial k_k^2 }\left[\left\langle\frac{1-\psi(u-i\bm k\cdot\bm v)}{u-i\bm k\cdot\bm v}\right\rangle\right.\nonumber \\&&\left.\times \frac{1}{1-\left\langle \psi(u-i\bm k\cdot\bm v)\right\rangle}\right],\nonumber
\end{eqnarray} 
where the RHS is taken at $\bm k=0$. We find using that odd moments of $v_i$, $v_k$ vanish that 
\begin{eqnarray}&& \!\!\!\!\!\!\!\!\!
\langle x_i^2(t)x_k^2(t)\rangle=\frac{\left\langle v_i^2v_k^2\right\rangle}{1-\psi(u)}\left(\frac{1-\psi(u)}{u}\right)^{(4)}+\frac{2\langle v_i^2\rangle\langle v_k^2\rangle \psi''(u)}{[1-\psi(u)]^2} \nonumber\\&&\!\!\!\!\!\!\!\!\!\times \left(\frac{1-\psi(u)}{u}\right)''+\frac{\left\langle v_i^2v_k^2\right\rangle \psi^{(4)}(u)}{u[1-\psi(u)]}+\frac{2\langle v_i^2\rangle\langle v_k^2\rangle [\psi''(u)]^2}{u[1-\psi(u)]^2}\nonumber.
\end{eqnarray} 
This can be used for detailed study of temporal behavior of $\langle x_i^2(t)x_k^2(t)\rangle$. We use this formula for finding the leading order long-time behavior in $XYZ...$ model. There we have $\left\langle v_i^2v_k^2\right\rangle=0$ for $i\neq k$, $\langle v_i^2\rangle=v_0^2/d$ which gives 
\begin{eqnarray}&& \!\!\!\!\!\!\!\!\!
\langle x_i^2(t)x_k^2(t)\rangle=\frac{2v_0^4 \psi''(u)}{[1-\psi(u)]^2d^2}\left[\left(\frac{1-\psi(u)}{u}\right)''+\frac{\psi''(u)}{u}\right]\nonumber.
\end{eqnarray} 
The leading order term when $u$ is small is 
\begin{eqnarray}&& \!\!\!\!\!\!\!\!\!
\langle x_i^2(t)x_k^2(t)\rangle=\frac{4v_0^4\alpha (\alpha-1)^2 A^2u^{2\alpha-7}}{\langle\tau\rangle^2d^2}.
\end{eqnarray} 
Performing inverse Laplace transform we find 
\begin{eqnarray}&& \!\!\!\!\!\!\!\!\!
\langle  x_i^2(t)x_k^2(t)\rangle=\frac{4v_0^4\alpha (\alpha-1)^2 A^2}{\Gamma(7-2\alpha)\langle\tau\rangle^2 d^2}t^{6-2\alpha}.
\end{eqnarray}
The probability distribution is characterized by constant time-independent ratio  
\begin{eqnarray}&& \!\!\!\!\!\!\!\!\!
\frac{\langle  x_i^2(t)x_k^2\rangle}{\langle  x_i^2(t)\rangle \langle x_k^2(t)\rangle}=\frac{4\alpha \Gamma^2(2-\alpha)}{\Gamma(7-2\alpha)},
\end{eqnarray}
telling that interdependence of the displacement's components becomes constant at large times. In contrast using $\left\langle v_i^4\right\rangle=v_0^4/d$ (this can be found from $v_0^4=\left\langle \left(\sum v_i^2\right)^2\right\rangle=d\langle v_i^4\rangle$ where we use that cross-correlations of velocity vanish) we find that the ratio 
\begin{eqnarray}&& \!\!\!\!\!\!\!\!\!
\frac{\langle x_i^2(t)x_k^2(t)\rangle}{\langle x_i^4(t)\rangle +\langle x_k^4(t)\rangle}=\frac{\alpha(\alpha-1) (4-\alpha)(5-\alpha) A\Gamma(2-\alpha)}{(2d)\langle\tau\rangle\Gamma(7-2\alpha)t^{\alpha-1}},\nonumber \end{eqnarray} decreases with time indefinitely characterizing growing anisotropy of the distribution.

\section{Infinite density is the generating function of the moments} \label{infinited}

We demonstrate that moments of integer order are described by the infinite density. We observe that Eqs.~(\ref{tr})-(\ref{tr1}) give for the Fourier transform of $I(\bm v)$ that,
\begin{eqnarray}&&\!\!\!\!\!\!\!\!\!\!\!\!\!\!
I(\bm k)=\frac{A}{|\Gamma(1-\alpha)|\langle\tau\rangle} \sum_{n=1}^{\infty}\frac{(2n)(-1)^n \left\langle (\bm k\!\cdot\!\bm v)^{2n}\right\rangle}{(2n)!(2n-\alpha)(2n+1-\alpha)}.\nonumber\end{eqnarray}
Comparing this with Eq.~(\ref{mom}) we conclude that in the limit of large times,
 \begin{eqnarray}&&\!\!\!\!\!\!\!\!\!\!\!\!\!\!
\frac{\langle x_{i_1}(t)x_{i_2}(t)\ldots x_{i_{2n}}(t)\rangle}{t^{2n+1-\alpha}} \sim \int v_{i_1}v_{i_2}\ldots v_{i_{2n}} I\left(\bm v\right) d\bm v. \nonumber
 \end{eqnarray}
This implies that, 
 \begin{eqnarray}&&\!\!\!\!\!\!\!\!\!\!\!\!\!\!
\langle x_{i_1}(t)x_{i_2}(t)\ldots x_{i_{2n}}(t)\rangle \sim \int \frac{x_{i_1}x_{i_2}\ldots x_{i_{2n}}}{t^{d+\alpha-1}} I\left(\frac{\bm x}{t}\right) d\bm x, \nonumber  
\end{eqnarray}
that is $P(\bm x, t)$ provided by Eq.~(\ref{infinite}) describes the long-time limit of the moments. In one-dimensional case this result was derived in \cite{BarkaiPhysRev}.

\section{Fractional derivative form of Fourier transform 
and isotropic L\'{e}vy distributions} \label{fraction}

In the recent work \cite{Marcin} numerical study of two-dimensional L\'{e}vy distributions was performed observing that the distribution can be written as fractional derivative of the distribution in one dimension and using the Matlab code for the fractional derivative. Here we observe that this consideration has universal applicablility. We demonstrate that $d-$dimensional Fourier transform of arbitrary radially symmetric function can be written as fractional derivative of order $(d-1)/2$ of the one-dimensional Fourier transform of that function. We clarify that this fact deserves to be known because it gives simple way of studying $d-$dimensional transforms based on simpler one-dimensional transform. Series expansions are obtained immediately using fractional derivatives of powers if term-by-term differentiation of series for one-dimensional Fourier transform is valid. Similarly the asymptotic form of the transform at large argument can be found by differentiation of simpler one-dimensional form. Numerically $d-$dimensional transforms are obtained applying fractional derivative code on well-developed one-dimensional Fourier transform code. 

We do the calculations for inverse Fourier transform with formulas for direct transform implied. We consider the $d-$dimensional Fourier transform of radially symmetric function $f_d(x)$, 
\begin{eqnarray}&&\!\!\!\!\!\!\!\!\!  
f(k)=\int \exp[-i\bm k\cdot\bm x]f_d(x)d\bm x,
\end{eqnarray}
where $|\bm x|=x$. The formula for the inverse Fourier transform of radially symmetric function whose Fourier transform is $f(k)$ where $k=|\bm k|$ is 
\begin{eqnarray}&&\!\!\!\!\!\!\!\!\!  
f_d(x)=\frac{x^{1-d/2}}{(2\pi)^{d/2}}\int_0^{\infty}J_{d/2-1}(kx)k^{d/2} f(k)dk.
\end{eqnarray}
We introduce $f_d(x)={\tilde f}_d(x^2)$ where 
\begin{eqnarray}&&\!\!\!\!\!\!\!\!\!  
{\tilde f}_d(x)=\frac{x^{1/2-d/4}}{(2\pi)^{d/2}}\int_0^{\infty}J_{d/2-1}(k\sqrt{x})k^{d/2} f(k)dk,\label{otld}
\end{eqnarray}
where $d$ is the dimension of space. We find the dependence of $f_d$ on $d$ when $f(k)$ is fixed function. We observe that operator of fractional derivative of order $1/2$ whose action on arbitrary well-behaved function $h$ obeys,  
\begin{eqnarray}&&\!\!\!\!\!\!\!\!\!  
D_-^{1/2} h=-\frac{d}{dx}\frac{1}{\sqrt{\pi}}\int_x^{\infty}\frac{h(x')dx'}{\sqrt{x'-x}},
\end{eqnarray}
is dimension raising, 
\begin{eqnarray}&&\!\!\!\!\!\!\!\!\!  
\frac{1}{\sqrt{\pi}}D_-^{1/2} {\tilde f}_d={\tilde f}_{d+1}. \label{raise}
\end{eqnarray}
We use that the identity, 
\begin{eqnarray}&&\!\!\!\!\!\!\!\!\!  
\frac{d}{dx} \left[x^{-\nu}J_{\nu}(x)\right]=-x^{-\nu}J_{\nu+1}(x),
\end{eqnarray}
implies 
\begin{eqnarray}&&\!\!\!\!\!\!\!\!\!  
\frac{d}{dx} \left[x^{-\nu/2}J_{\nu}(k\sqrt{x})\right]=-k\frac{x^{-(\nu+1)/2}J_{\nu+1}(k\sqrt{x})}{2}.
\end{eqnarray}
We observe that $D_-^{1/2}$ can be written in terms of the right-side fractional integral $I_-^{1/2}$ defined as \cite{fr},
\begin{eqnarray}&&\!\!\!\!\!\!\!\!\!  
I_-^{1/2} [h]=\frac{1}{\sqrt{\pi}}\int_x^{\infty}\frac{h(x')dx'}{\sqrt{x'-x}}. \label{dfn}
\end{eqnarray}
We have
\begin{eqnarray}&&\!\!\!\!\!\!\!\!\!
D_-^{1/2}[h]=-I_-^{1/2}[h'],\label{in} 
\end{eqnarray}
where we use that,
\begin{eqnarray}&&\!\!\!\!\!\!\!\!\!  
\frac{d}{dx}\int_x^{\infty}\frac{h(x')dx'}{\sqrt{x'-x}}=2\frac{d}{dx}\int_x^{\infty} h(x')dx'\frac{d}{dx'}\sqrt{x'-x}\nonumber\\&&\!\!\!\!\!\!\!\!\!
=-2\frac{d}{dx}\int_x^{\infty} h'(x')dx'\sqrt{x'-x}=\int_x^{\infty}\frac{h'(x')dx'}{\sqrt{x'-x}}.\label{inta}
\end{eqnarray}
We find using fractional integral from \cite{bat} that,
\begin{eqnarray}&&\!\!\!\!\!\!\!\!\!  
\frac{1}{\sqrt{\pi}}D_-^{1/2} \left[x^{-\nu/2}J_{\nu}(k\sqrt{x})\right]=\frac{k}{2\pi}\int_x^{\infty}\frac{x'^{-(\nu+1)/2}dx'}{\sqrt{x'-x}}\nonumber\\&&\!\!\!\!\!\!\!\!\!\times J_{\nu+1}(k\sqrt{x'}) =\sqrt{\frac{k}{2\pi}} x^{-(\nu+1/2)/2}J_{\nu+1/2}(k\sqrt{x}). \label{id}
\end{eqnarray}
Thus we find Eq.~(\ref{raise}) by acting on Eq.~(\ref{otld}) with $D_-^{1/2}$ and using identity (\ref{id}). Further applying $d-1$ times $D_-^{1/2}$ on ${\tilde f}_1$ and using Eq.~(\ref{raise}) we obtain that, 
\begin{eqnarray}&&\!\!\!\!\!\!\!\!\!  
\frac{1}{\pi^{(d-1)/2}}\left[D_-^{1/2}\right]^{d-1} {\tilde f}_1={\tilde f}_{d}.\label{it}
\end{eqnarray}
If $D_-^{1/2}$ would be ordinary derivative then the above would imply 
\begin{eqnarray}&&\!\!\!\!\!\!\!\!\!  
\frac{1}{\pi^{(d-1)/2}}D_-^{(d-1)/2} {\tilde f}_1={\tilde f}_{d},\label{dm}
\end{eqnarray}
where the fractional derivative of order $\alpha$ is,
\begin{eqnarray}&&\!\!\!\!\!\!\!\!\!  
D_-^{\alpha}f=\frac{(-1)^n}{\Gamma(n-\alpha)}\frac{d^n}{dx^n}\int_{x}^{\infty} \frac{f(x') dx'}{(x'-x)^{\alpha-n+1}},
\end{eqnarray} 
\end{appendices} 
\begin{thebibliography}{99}

\bibitem{Sinai} Ya. G. Sinai, {\it Introduction to Ergodic Theory}, (Princeton University Press, 1976).

\bibitem{bunim}  L. A. Bunimovich, et al., {\it Hard Ball Systems and the Lorentz Gas}, (Springer Science and Business Media, 2013).

\bibitem{mw}  E. Montroll and G. Weiss, J. Math. Phys. \textbf{6}, 167 (1965).

\bibitem{sl} H. Scher and M. Lax, Phys. Rev. B 7, 4491 (1973).

\bibitem{Shl} M. Shlesinger, J. Stat. Phys. \textbf{10}, 421 (1974).

\bibitem{Mon} H. Scher and E. W. Montroll, Phys. Rev. B \textbf{12}, 2455 (1975).

\bibitem{Shlesinger82} M. F. Shlesinger, J. Klafter, and Y. M. Wong, J. Stat. Phys. \textbf{27}, 499 (1982).

\bibitem{bg} J. P. Bouchaud and A. Georges, Phys. Rep. \textbf{195}, 127 (1990).

\bibitem{Isichenko} M. B. Isichenko, Rev. Mod. Phys. \textbf{64}, 9611043(1992).

\bibitem{Shlesinger93} M. F. Shlesinger, G. M. Zaslavsky, and J. Klafter, Nature \textbf{363}, 3137(1993).

\bibitem{Shlesinger96}  J. Klafter, M. F. Shlesinger, and G. Zumofen, Phys. Today, 33(1996).

\bibitem{UchZol} V. V. Uchaikin and V. M. Zolotarev, {\it Chance and Stability: Stable Distributions and their Applications} (Walter de Gruyter, 1999).

\bibitem{km} R. Metzler and J. Klafter, Phys. Rep. \textbf{339}, 1 (2000),
ISSN 0370-1573.

\bibitem{godreche} C. Godreche and J. M. Luck, J. Stat. Phys., \textbf{104}, 489 (2001). 

\bibitem{ks} J. Klafter and I. Sokolov, Phys. World \textbf{18}, 29 (2005).

\bibitem{xper} N. Gal and D. Weihs, Phys. Rev. E, \textbf{81}, 020903 (2010).

\bibitem{BarkaiPhysRev}  A. Rebenshtok, S. Denisov, P. H\"{a}nggi, and E. Barkai, Phys. Rev. E \textbf {90}, 062135 (2014).

\bibitem{BarkaiPhysLett}  A. Rebenshtok, S. Denisov, P. H\"{a}nggi, and E. Barkai, Phys. Rev. Lett. \textbf{112}, 110601 (2014).

\bibitem{BarkaiPR} D. Froemberg, M. Schmiedeberg, E. Barkai, and V. Zaburdaev, Phys. Rev. E \textbf{91}, 022131 (2015).

\bibitem{Zaburdaev} V. Zaburdaev, S. Denisov, and J. Klafter, Rev. Mod. Phys., \textbf{87}, 483 (2015).

\bibitem{Marcin} M. Magdziarz and T. Zorawik, arxiv, 1510.05614.

\bibitem{recent} V. Zaburdaev, I. Fouxon, S. Denisov, and E. Barkai, arXiv, 1605.02908.

\bibitem{arxiv} J. P. Taylor-King, R. Klages, S. Fedotov, and R. A. Van Gorder, Phys. Rev. E \textbf{94}, 012104 (2016). 

\bibitem{kb} D. A. Kessler and E. Barkai, Phys. Rev. Lett. \textbf{108}, 230602 (2012).

\bibitem{Burioni} R. Burioni, L. Caniparoli, and A. Vezzani, Phys. Rev. E, \textbf{81}, 060101 (2010).

\bibitem{Dentz} M. Dentz, T. Le Borgne, D. R. Lester, and F. P. de Barros, Phys. Rev. E, \textbf{92}, 032128 (2015). 

\bibitem{Agliari} E. Agliari and R. Burioni, Phys. Rev. E, \textbf{80}, 031125 (2009).

\bibitem{bd} J. P. Bouchaud and P. Le Doussal, J. Stat. Phys. \textbf{41}, 225 (1985).

\bibitem{cristadoro} G. Cristadoro, Th. Gilbert, M. Lenci, D. P. Sanders, Phys. Rev. E 90, 050102 (2014).

\bibitem{den1} I. Fouxon, S. Denisov, E. Barkai, \textit{to be published}.

\bibitem{note1} The regime with infinite mean step time is distinctive even in the one-dimensional case; 
see \cite{BarkaiPhysRev,BarkaiPhysLett,BarkaiPR,Marcin}.

\bibitem{antipod} Ch. Bingham,  Ann. Stat. 2, 1201 (1974).

\bibitem{direction} K. V. Mardia and P. Jupp, \textit{Directional Statistics} (John Wiley and Sons, 2000).

\bibitem{Feller} W. Feller, {\it An Introduction to Probability Theory}, vols. 1 and 2, (Wiley, New York, 1971).

\bibitem{ellis1} R. S. Ellis, {\it Entropy, Large Deviations, and Statistical Mechanics}, vol. 271, (Springer Science and Business Media, 2012).

\bibitem{ellis2} R. S. Ellis, Physica D, \textbf{133}, 106 (1999).

\bibitem{fb} E. Balkovsky and A. Fouxon, Phys. Rev. E \textbf{60}, 4164 (1999).

\bibitem{Rebenshtok} A. Rebenshtok, S. Denisov, P. H\"{a}nggi, and E. Barkai, Math. Model. Nat. Phenom. \textbf{11}, 76 (2016).
\bibitem{lv} P. L\'{e}vy, {\it Theorie de l'addition des variables aleatoires}, (Gauthiers-Villars, Paris, 1937).

\bibitem{kg}  B. V. Gnedenko and A. N. Kolmogorov, Amer. J. Math. \textbf{105}, 28 (1954).

\bibitem{st} G. Samorodnitsky and M. S. Taqqu, {\it Stable Non–Gaussian Random Processes}, (Chapman and Hall, New York, 1994).

\bibitem{levy} P. L\'{e}vy, {\it Calcul des Probabilites}, (Gauthier Villars, Paris, 1925).

\bibitem{Itzykson} C. Itzykson and D. Drouffe, {\it Statistical Field Theory: Volume 1, From Brownian Motion to Renormalization and Lattice Gauge Theory}, (Cambridge University Press, Cambridge, UK, 1989). 

\bibitem{aa} J. Aaronson, {\it An Introduction to Infinite Ergodic Theory}, (American Mathematical Society, Providence, RI, 1997).

\bibitem{tz} M. Thaler and R. Zweimuller, Probab. Theor. Relat. Fields \textbf{135}, 15 (2006).
 

\bibitem{psd} O. Arizmendi and V. Pérez-Abreu, Commun. Stoch. Anal \textbf{4}, 161 (2010).

\bibitem{nebr} P. Nebres,  {\it Renewal theory and its applications}, 2011. 
\bibitem{fr} A. A. Kilbas, H. M. Srivastava, and J. J. Trujillo, {\it Theory and Applications of Fractional Differential Equations}, (North - Holland Mathematics Studies 204, Amsterdam, 2006).

\bibitem{bat} H. Bateman, {\it Higher Transcendantal Functions},  vol. $2$ (New York: McGraw-Hill, 1955).











\end{thebibliography}
\end{document}